\def\ph{\phantom{-}}  
\def\mb#1{\mbox{\boldmath{$#1$}}}
\def\sech{\rm sech}
\documentclass[amsmath,amssymb,superscriptaddress,showkeys, showpacs, aps]{revtex4}
\usepackage{graphicx}
\usepackage{units}
\usepackage{mathrsfs} 
\usepackage{amssymb}
\usepackage{cases}
\makeatletter
\newcommand*{\rom}[1]{\expandafter\@slowromancap\romannumeral #1@}
\makeatletter
\newcommand{\rmn}[1]{\romannumeral #1}
\newcommand{\Rmn}[1]{\expandafter\@slowromancap\romannumeral #1@}
\makeatother

\begin{document}

\hspace*{5.0 in}CUQM - 153
%\hspace*{5.0 in}[drcom {$22^{\rm nd}$} April 2015]
\vspace{0.5 in}
%\markboth{R.~L.~Hall \& P. Zorin}{Refined comparison theorems for the Dirac equation.}

%%%%%%%%%%%%%%%%%%%%%%%%%%%%%%%%%%%%%%%%%%%%%%%%%%%%%%%%%%%%%%%%%%%%%%%%%%%%
\title{Refined comparison theorems for the Dirac equation in $d$ dimensions.}
%%%%%%%%%%%%%%%%%%%%%%%%%%%%%%%%%%%%%%%%%%%%%%%%%%%%%%%%%%%%%%%%%%%%%%%%%%
\author{Richard L. Hall}
\email{richard.hall@concordia.ca}
\affiliation{Department of Mathematics and Statistics, Concordia University,
1455 de Maisonneuve Boulevard West, Montr\'eal,
Qu\'ebec, Canada H3G 1M8}

\author{Petr~Zorin}
\email{petrzorin@yahoo.com}
\affiliation{Department of Mathematics and Statistics, Concordia University,
1455 de Maisonneuve Boulevard West, Montr\'eal,
Qu\'ebec, Canada H3G 1M8}

%%%%%%%%%%%%%%%%%%%%%%%%%%%%%%%%%%%%%%%%%%%%%%%%%%%%%%%%%%%%%%%%%%%%%%%%%%%
%%%%%%%%%%%%%%%%%%%%%%%%%%%%%%%%%%%%%%%%%%%%%%%%%%%%%%%%%%%%%%%%%%%%%%%%%%%
\begin{abstract} 
A single spin-$\nicefrac{1}{2}$ particle obeys the Dirac equation in $d\ge 1$  spatial dimension and is bound by an attractive central monotone potential which vanishes at infinity (in one dimension the potential is even). This work refines the relativistic comparison theorems which were derived by Hall \cite{p75}. The new theorems allow the graphs of the two comparison potentials $V_a$ and $V_b$ to crossover in a controlled way and still imply the spectral ordering $E_a\le E_b$ for the eigenvalues at the bottom of each angular momentum subspace. More specifically in a simplest case we have: in dimension $d=1$, if $\int_0^x (V_b(t)-V_a(t)) dt\ge 0,\ x\in [0,\ \infty)$, then $E_a\le E_b$; and in $d>1$ dimensions, if $\int_0^r (V_b(t)-V_a(t))t^{2|k_d|} dt\ge 0,\ r\in [0,\ \infty)$, where $k_d=\tau\left(j+\frac{d-2}{2}\right)$ and $\tau=\pm 1$, then $E_a\le E_b$. 
\end{abstract}
%%%%%%%%%%%%%%%%%%%%%%%%%%%%%%%%%%%%%%%%%%%%%%%%%%%%%%%%%%%%%%%%%%%%%%%%%%%
%%%%%%%%%%%%%%%%%%%%%%%%%%%%%%%%%%%%%%%%%%%%%%%%%%%%%%%%%%%%%%%%%%%%%%%%%%%



\maketitle

%%%%%%%%%%%%%%%%%%%%%%%%%%%%%%%%%%%%%%%%%%%%%%%%%%%%%%%%%%%%%%%%%%%%%%%%%%%%%%%%%%%%%%%%
%%%%%%%%%%%%%%%%%%%%%%%%%%%%%%%%%%%%%%%%%%%%%%%%%%%%%%%%%%%%%%%%%%%%%%%%%%%%%%%%%%%%%%%%
\section{Introduction}
%%%%%%%%%%%%%%%%%%%%%%%%%%%%%%%%%%%%%%%%%%%%%%%%%%%%%%%%%%%%%%%%%%%%%%%%%%%%%%%%%%%%%%%%
%%%%%%%%%%%%%%%%%%%%%%%%%%%%%%%%%%%%%%%%%%%%%%%%%%%%%%%%%%%%%%%%%%%%%%%%%%%%%%%%%%%%%%%%
The comparison theorem of quantum mechanics states that if two comparison potentials are ordered, i.e. $V_a\le V_b$, then the discrete energy eigenvalues are ordered as well $E_a\le E_b$. In the nonrelativistic case this is a straightforward consequence of the min--max variational characterization of the discrete part of the spectrum \cite{Reed, Thirring}. In the relativistic case the Hamiltonian is not bounded below, and a variational analysis is more complicated \cite{Fr, Gold, Gr}. However, comparison theorems have been established by other means in $d=1$ and $d=3$ dimensions \cite{p75}, in $d=2$ dimensions \cite{chen1}, and in $d$ dimensions \cite{chen2}, most recently by monotonicity arguments \cite{p127, monoton2, HY, p134}. In Ref. \cite{Semay} Semay used the Hellmann--Feynman theorem \cite{HF} to established a general comparison theorem for the  Schr\"{o}dinger and Dirac equations.\smallskip

In this paper we derive {\it refined} comparison theorems which allow the graphs of the comparison potentials to cross over in a controlled fashion and still imply definite ordering of the respective eigenvalues at the bottom of each angular-momentum subspace. This idea was first explored by Hall {\it et al} for nonrelativistic problems in $d=1$ and $d=3$ dimensions \cite{Hall7} and in $d>1$ dimensions \cite{ddimSch}, and applied to Sturm--Liouville problems in \cite{Hall44}. In the simplest case one derives the spectral ordering $E_a\le E_b$ from the weaker potential assumption $U_a\le U_b$, where $U_i=\int_0^x V_i(t)dt$, $i=a$ or $b$. Since these refined nonrelativistic results were obtained {\it without} the use of a variational characterization of the discrete spectrum, similar reasoning could be applied to derive a basic relativistic comparison theorem for the Dirac equation \cite{p75}. The principal aim of the present paper is to go further and derive refined comparison theorems also for the Dirac spectral problem itself. \smallskip

In dimension $d=1,$ the energies compared are simply the lowest discrete eigenvalues.  In $d>1$ dimensions, the energies are the lowest eigenvalues in each angular-momentum sector.  The derivations rely on {\it a priori} knowledge of the nodal structure characterized for central fields in Refs.\,\cite{rose, rose_book,nod}.  We found it necessary  to discuss the cases  $ d=1$ and $d>1$ separately and to treat a small number of distinct classes of attractive monotone potentials.  Sharper energy bounds can be obtained if the component wave functions are also known for the chosen base comparison potential.
 Simple sufficient conditions are derived in corollaries to the comparison theorems to make their use more immediate and straightforward. The results are illustrated by some specific examples.

%%%%%%%%%%%%%%%%%%%%%%%%%%%%%%%%%%%%%%%%%%%%%%%%%%%%%%%%%%%%%%%%%%%%%%%%%%%%%%%%%%%%%%%%
%%%%%%%%%%%%%%%%%%%%%%%%%%%%%%%%%%%%%%%%%%%%%%%%%%%%%%%%%%%%%%%%%%%%%%%%%%%%%%%%%%%%%%%%
\section{Dirac equation in one dimension}
%%%%%%%%%%%%%%%%%%%%%%%%%%%%%%%%%%%%%%%%%%%%%%%%%%%%%%%%%%%%%%%%%%%%%%%%%%%%%%%%%%%%%%%%
%%%%%%%%%%%%%%%%%%%%%%%%%%%%%%%%%%%%%%%%%%%%%%%%%%%%%%%%%%%%%%%%%%%%%%%%%%%%%%%%%%%%%%%%
The Dirac equation in one spatial dimension for a single spin-$\nicefrac{1}{2}$ particle of mass $m$ in natural units $\hbar=c=1$ may be written \cite{calog}: 
\begin{equation*}
\left(\sigma_1\frac{\partial}{\partial x}-(E-V)\sigma_3+m\right)\psi=0, 
\end{equation*}
where $\sigma_1$ and $\sigma_3$ are Pauli matrices and the discrete energy eigenvalue $E$ such that $-m<E<m$, \cite{Spectrumd11, Spectrumd12}. 
The vector potential $V$ (the time component of a four--vector) satisfies
\begin{eqnarray*}
&(i)&V\ \text{is even, i.e.} \  V(x)=V(-x);\\
\vspace{9mm}
&(ii)&V\ \text{is nonpositive and bounded, i.e.} \ V_0\le V\le 0,\ \text{where}\ V_0=V(0);\\
\vspace{9mm}
&(iii)&V\ \text{vanishes at infinity, thus}\ \lim_{x\to\pm\infty}V=0;\\
\vspace{9mm}
&(iv)&V\ \text{is attractive, that is monotone nondecreasing on} \  [0, \infty).
\end{eqnarray*}
By taking the two--component Dirac spinor as $\psi=\left(\begin{array}{cc}\varphi_1 \\ \varphi_2\end{array}\right)$ the above matrix equation can be decomposed into a system of first--order linear differential equations \cite{Dombey, Qiong}:
\begin{subnumcases}{}
\label{d1l}
\varphi_1'=-(E+m-V)\varphi_2,\\
\label{d1r}
\varphi_2'=\ph(E-m-V)\varphi_1,
\end{subnumcases}
where prime $'$ denotes the derivative with respect to $x$. For bound states, $\varphi_1$ and $\varphi_2$ satisfy the normalization condition
\begin{equation*}
(\varphi_1,\varphi_1) + (\varphi_2,\varphi_2) = \int\limits_{-\infty}^{\infty}(\varphi_1^2 + \varphi_2^2)dx = 1.
\end{equation*}

For the reason which will be clear later, the comparison theorem below requires knowledge concerning the ground state. According to the Nodal Theorem of Ref.\cite{nod}, the upper and lower components of the Dirac spinor, $\varphi_1$ and $\varphi_2$ respectively, have definite and opposite parities and $n_2=n_1+1$, where $n_i$, $i=1$ or $2$, the corresponding number of nodes of $\varphi_i$. Thus in the state with the smallest number of nodes, the upper component $\varphi_1$ is even and the lower one $\varphi_2$ is odd. From now on, without loss of generality, we consider the interval $[0, \infty)$ and assume that both components of the Dirac spinor lie above the $x$--axis, i.e. $\varphi_1 \ge 0$ and $\varphi_2 \ge 0$ on $[0, \infty)$. Then it follows from (\ref{d1l})--(\ref{d1r}) that on $[0, \infty)$, $\varphi_1'\le 0$ and $\varphi_2'\ge 0$ near the origin then $\varphi_2'\le 0$ at infinity. As an illustration we plot $\{\varphi_1, \varphi_2\}$ for the exponential potential $V=-\beta e^{-b|x|}$ \cite{Flugge} (see Figure~1). 
\begin{figure}
\centering{\includegraphics[height=13cm,width=5cm,angle=-90]{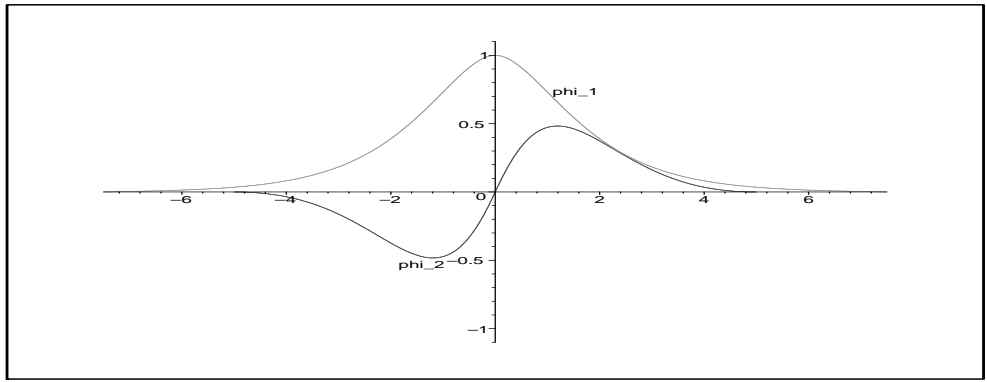}}
\caption{Ground state wave functions $\varphi_1$ and $\varphi_2$, corresponding to the exponential potential $V=-\beta e^{-b|x|}$, with $\beta=0.9$, $b=0.5$, $m=1$, and energy is $E=0.49233$.}
\end{figure}

%%%%%%%%%%%%%%%%%%%%%%%%%%%%%%%%%%%%%%%%%%%%%%%%%%%%%%%%%%%%%%%%%%%%%%%%%%%%%%%%%%%%%%%%
%%%%%%%%%%%%%%%%%%%%%%%%%%%%%%%%%%%%%%%%%%%%%%%%%%%%%%%%%%%%%%%%%%%%%%%%%%%%%%%%%%%%%%%%
\section{Refined comparison theorems for the Dirac equation in one dimension}
%%%%%%%%%%%%%%%%%%%%%%%%%%%%%%%%%%%%%%%%%%%%%%%%%%%%%%%%%%%%%%%%%%%%%%%%%%%%%%%%%%%%%%%%
%%%%%%%%%%%%%%%%%%%%%%%%%%%%%%%%%%%%%%%%%%%%%%%%%%%%%%%%%%%%%%%%%%%%%%%%%%%%%%%%%%%%%%%%
We compare two problems with symmetric potentials $V_a$ and $V_b$ and ground state energies $E_a$ and $E_b$ for which the system (\ref{d1l})--(\ref{d1r}) becomes respectively
\begin{subnumcases}{}
\label{3}
\varphi_{1a}'=-(E_a+m-V_a)\varphi_{2a},\\
\label{4}
\varphi_{2a}'=\ph(E_a-m-V_a)\varphi_{1a},
\end{subnumcases}
and
\begin{subnumcases}{}
\label{5}
\varphi_{1b}'=-(E_b+m-V_b)\varphi_{2b},\\
\label{6}
\varphi_{2b}'=\ph(E_b-m-V_b)\varphi_{1b}.
\end{subnumcases}
Let us consider the combination of equations: 
\begin{equation}\label{comb}
(\ref{3})\varphi_{2b} - (\ref{4})\varphi_{1b} - (\ref{5})\varphi_{2a} + (\ref{6})\varphi_{1a},
\end{equation}
which after some simplifications becomes
\begin{equation*}
(\varphi_{1a}\varphi_{2b})'-(\varphi_{2a}\varphi_{1b})'=(\varphi_{1a}\varphi_{1b}+
\varphi_{2a}\varphi_{2b})\left[(E_b-E_a)-(V_b-V_a)\right].
\end{equation*}
Integrating the left side of the above expression by parts from $0$ to $\infty$ and using the boundary conditions, $\varphi_{1a}(0)=\varphi_{1b}(0)=0$ and $\lim\limits_{x\to\infty}\varphi_{2a}=\lim\limits_
{x\to\infty}\varphi_{2b}=0$, we find $\int_0^\infty \left[(\varphi_{1a}\varphi_{2b})'-(\varphi_{2a}\varphi_{1b})'\right]dx=0$. 
Then we integrate the right side to obtain
\begin{equation}\label{9}
(E_b-E_a)\int_0^\infty (\varphi_{1a}\varphi_{1b}+\varphi_{2a}\varphi_{2b})dx=\int_0^\infty (\varphi_{1a}\varphi_{1b}+\varphi_{2a}\varphi_{2b})(V_b-V_a)dx.
\end{equation}
It follows from the last expression that if the wave functions have no nodes, so that the integrands have constant signs, and the potentials are ordered i.e. $V_a\le V_b$, then $E_a\le E_b$. This is the comparison theorem which was first proved in $d=3$ dimensions in \cite{p75}, in $d=2$ dimensions in \cite{chen1}, and in $d$ dimensions in \cite{chen2}. Later, using the monotonicity concept, the comparison theorem was proved for higher dimension cases $d\ge 1$ for the Dirac equation in \cite{p127} and the Klein--Gordon equation in \cite{p134} for all the excited states. 

The Dirac equation admits exact analytical solutions for very few potentials. The above theorem allows us to obtain upper or lower bounds for any eigenvalue with the aid of suitable comparison potentials. But the comparison potentials can not cross each other, because in that case the integrands of (\ref{9}) change sign. Similarly to the nonrelativistic case \cite{Hall7} we now derive refined relativistic comparison theorems which allow the graphs of the potentials to crossover in a controlled manner so that spectral ordering is predicted. 

\medskip

\noindent{\bf Theorem 1:} ~~{\it The potential $V$  satisfies $(i)$--$(iii)$, $-V_0\le 2m$, and has area. Then if 
\begin{equation}\label{theorem1}
g(x)=\int_0^x (V_b(t)-V_a(t))dt\ge 0, \quad  x\in [0,\ \infty),
\end{equation}
we have $E_a\le E_b$.} 

\medskip

\noindent{\bf Proof:} We integrate the right side of (\ref{9}) by parts to obtain
\begin{equation*}
\int_0^\infty (\varphi_{1a}\varphi_{1b}+\varphi_{2a}\varphi_{2b})(V_b-V_a)dx=
(\varphi_{1a}\varphi_{1b}+\varphi_{2a}\varphi_{2b})g|_0^\infty - \int_0^\infty g(\varphi_{1a}\varphi_{1b}+\varphi_{2a}\varphi_{2b})'dx,
\end{equation*}
where $g(x)$ is defined by (\ref{theorem1}). Since $g(0)=0$ and $\lim\limits_{x\to\infty}\varphi_{1a}=\lim\limits_{x\to\infty}\varphi_{2a}=0$, relation (\ref{9}) becomes
\begin{equation}\label{expr4}
(E_b-E_a)\int_0^\infty (\varphi_{1a}\varphi_{1b}+\varphi_{2a}\varphi_{2b})dx=
- \int_0^\infty g(\varphi_{1a}\varphi_{1b}+\varphi_{2a}\varphi_{2b})'dx.
\end{equation}

In order to find the sign of $(\varphi_{1a}\varphi_{1b}+\varphi_{2a}\varphi_{2b})'$
we write
\begin{equation*}
(\varphi_{1a}\varphi_{1b}+\varphi_{2a}\varphi_{2b})'=W_1\varphi_{1b}\varphi_{2a}
+W_2\varphi_{1a}\varphi_{2b},
\end{equation*}
where
\begin{equation*}\label{W1}
W_1=-E_a+E_b-2m+V_a-V_b \quad \text{and} \quad 
W_2=E_a-E_b-2m-V_a+V_b.
\end{equation*}
Let $\varphi_2$ represent either $\varphi_{2a}$ or $\varphi_{2b}$. Suppose that $\varphi_2$ reaches its maximum at some point $x_c$; thus $\varphi_2'\ge 0$ on $[0, x_c]$ and $\varphi_2'\le 0$ on $[x_c, \infty)$. It follows from (\ref{d1r}) that $E_a-m-V_a\ge 0$ on $[0, x_c]$, so $W_1<E_b-V_b-3m$. Since $-V_b\le 2m$ and $-m<E_b<m$, $W_1<0$ on $[0,\ x_c]$. Then on $[x_c, \infty)$ equation (\ref{d1r}) implies $E_b-m-V_b\le 0$, which leads to $W_1\le -E_a-m+V_a$, energy $E_a$ satisfies $-m<E_a<m$ so $W_1< 0$ on $[x_c, \infty)$. Therefore $W_1<0$ on $[0, \infty)$. Similarly it can be shown that $W_2<0$ on $[0, \infty)$. Thus $(\varphi_{1a}\varphi_{1b}+\varphi_{2a}\varphi_{2b})'\le 0$, on $[0, \infty)$. 
Therefore if $g(x)\ge 0$ relation (\ref{expr4}) ensures that $E_a\le E_b$, which result completes the proof of the theorem.

\hfill $\Box$ 

\medskip

If we know the exact behaviour of the comparison potentials we can state simpler sufficient conditions for spectral ordering: 

\medskip

\noindent{\bf Corollary 1:} ~~{\it If the potentials cross over once, say at $x_1$, $V_a\le V_b$ for $x\in [0,\ x_1]$, and
\begin{equation*}\label{concl2}
g(\infty)=\int_0^\infty (V_b-V_a)dx\ge 0,  
\end{equation*}
then $E_a\le E_b$. If the potentials cross over twice, say at $x_1$ and $x_2$, $x_1<x_2$, $V_a\le V_b$ for $x\in [0,\ x_1]$, and
\begin{equation*}\label{concl1}
g(x_2)=\int_0^{x_2} (V_b-V_a) dx\ge 0, 
\end{equation*}
then $E_a\le E_b$.} 

\medskip

We note that such application of Theorem 1 via the Corollary 1 can easily be extended. For example, consider the case of $n$ intersections, $n=1,\ 2,\ 3,\ \ldots$, and suppose again that $V_a\le V_b$ for the first interval $x\in[0,\ x_1]$. Suppose now that the sequence $\int_{x_i}^{x_{i+1}}|V_b-V_a|dx$, $i=1,\ 2,\ 3,\ \ldots,\ n$, of absolute areas is nonincreasing (if $n$ is odd then $\int_{x_{n-1}}^{x_n}|V_b-V_a|dx\ge\int_{x_n}^\infty|V_b-V_a|dx$ ), consequently it follows that $\int_0^x (V_b(t)-V_a(t))dt\ge 0$, $x\in[0,\ \infty)$, and we conclude $E_a\le E_b$.
\smallskip

\noindent {\bf Remark:} we now also consider theorems which take advantage of the known wave functions for {\it one} of the two comparison potentials. The general concept here is that we use these known wave functions for one of the eigenproblems along with an assumed relationship between the two potentials, and from these conditions we predict bounds on the eigenvalues of the second problem. In each such theorem we choose the base comparison potential to be $V_i$ where in an application $i$ may be chosen to be either $i=a$ or $i=b$; of course, changing the base problem will 
also reverse the energy inequality from lower to upper bound, or {\it vice versa}.
\smallskip
Now we state the second theorem (which allows the bottom of the potential to lie below $-2m$):

\medskip

\noindent{\bf Theorem 2:} ~~{\it The potential $V$  satisfies $(i)$--$(iv)$ and has $\varphi_{1i}$ and $t\varphi_{2i}$--weighted areas, if 
\begin{equation}\label{claim2}
k_1(x)=\int_0^x (V_b(t)-V_a(t))\varphi_{1i}(t)dt\ge 0 \quad \text{and} \quad
k_2(x)=\int_0^x (V_b(t)-V_a(t))\varphi_{2i}(t)tdt\ge 0, 
\quad x\in [0,\ \infty),
\end{equation}
where $i$ is either $a$ or $b$, then we have $E_a\le E_b$.} 

\medskip

\noindent{\bf Proof:} We prove the theorem for $i=b$; for $i=a$, the proof is the same. We integrate the right side of (\ref{9}) by parts to obtain 
\begin{equation*}
(E_b-E_a)\int_0^\infty (\varphi_{1a}\varphi_{1b}+\varphi_{2a}\varphi_{2b})dx=
\left[k_1\varphi_{1a}+k_2\frac{\varphi_{2a}}{x}\right]_0^\infty
-\int_0^\infty \left[k_1\varphi'_{1a} + k_2\left(\frac{\varphi_{2a}}{x}\right)'\right]dx,
\end{equation*}
where $k_1(x)$ and $k_2(x)$ are defined by (\ref{claim2}) for $i=b$. The expression $\left[k_1\varphi_{1a}+k_2\frac{\varphi_{2a}}{x}\right]_0^\infty=0$, because $k_1(0)=k_2(0)=0$ and $\lim\limits_{x\to\infty}\varphi_{1a}=\lim\limits_{x\to\infty}\varphi_{2a}=0$.  
The function $\varphi_{1a}$ is positive and decreasing, thus $\varphi'_{1a}\le 0$ on $[0, \infty)$. We know that $\varphi'_{2a}\ge 0$ on $[0, x_c]$ and $\varphi_{2a}'\le 0$ on $[x_c, \infty)$. It follows that $\left(\varphi_{2a}/x\right)'\le 0$ on $[x_c, \infty)$. From the assumption $(iv)$ and  (\ref{4}) we conclude $\varphi''_{2a}\le 0$, that is to say $\varphi_{2a}$ is concave on $[0, x_c]$, so lies below its tangents lines: thus  $\varphi_{2a}(x)\ge x\varphi_{2a}'(x)$, which implies $\left(\varphi_{2a}/x\right)'< 0$ on $[0, x_c]$. Therefore $\left(\varphi_{2a}/x\right)'\le 0$ on $[0, \infty)$. Finally, if $k_1(x)$ and $k_2(x)$ are both nonnegative it follows from the above expression that $E_a\le E_b$. This completes the proof.

\hfill $\Box$

\medskip

The second theorem is sronger because the potential difference $\triangle V = V_b - V_a$ is multiplied by the decreasing factor $\varphi_{1i}$ in $k_1(x)$, and by $t\varphi_{2i}$ in $k_2(x)$, $i=a,\ b$, and this allows $\triangle V$ to be even larger than in Theorem 1 and still imply the spectral ordering $E_a\le E_b$. Similarly to Corollary 1, but now with the $\varphi_{ji}$--weighted areas, $j=1$ or $2$ and $i=a$ or $b$, we can state the following sufficient condition for spectral ordering:

\medskip

\noindent{\bf Corollary 2:} ~~{\it If the potentials cross over once, say at $x_1$, $V_a\le V_b$ for $x\in [0,\ x_1]$, and
\begin{equation*}\label{concl3}
k_1(\infty)=\int_0^\infty (V_b-V_a)\varphi_{1i}dx\ge 0 \quad \text{and} \quad
k_2(\infty)=\int_0^\infty (V_b-V_a)\varphi_{2i}xdx\ge 0, 
\quad i=a\ or\ b,
\end{equation*}
then $E_a\le E_b$. If the potentials cross over twice, say at $x_1$ and $x_2$, $x_1<x_2$, $V_a\le V_b$ for $x\in [0,\ x_1]$, and
\begin{equation*}\label{concl4}
k_1(x_2)=\int_0^{x_2} (V_b-V_a)\varphi_{1i}dx\ge 0 \quad \text{and} \quad
k_2(x_2)=\int_0^{x_2} (V_b-V_a)\varphi_{2i}xdx\ge 0, 
\quad i=a\ or\ b,  
\end{equation*}
then $E_a\le E_b$.} 

\medskip

Corollary 2 can be generalized as well for the case of $n$ intersections: if $V_a\le V_b$ on $x\in[0,\ x_1]$ and the sequences $\int_{x_i}^{x_{i+1}}|V_b-V_a|\varphi_{1j}dx$ and $\int_{x_i}^{x_{i+1}}|V_b-V_a|\varphi_{2j}xdx$, $i=1,\ 2,\ 3,\ \ldots,\ n$ and $j=a$ or $b$, of absolute areas are nonincreasing (and, if $n$ is odd, $\int_{x_{n-1}}^{x_n}|V_b-V_a|\varphi_{1j}dx\ge\int_{x_n}^\infty|V_b-
V_a|\varphi_{1j}dx$ and $\int_{x_{n-1}}^{x_n}|V_b-V_a|\varphi_{2j}xdx
\ge\int_{x_n}^\infty|V_b-V_a|\varphi_{2j}xdx$), then $\int_0^x (V_b(t)-V_a(t))\varphi_{1j}(t)dt\ge 0$ and $\int_0^x (V_b(t)-V_a(t))\varphi_{2j}(t)tdt\ge 0$ on $x\in[0,\ \infty)$, so, according to Theorem 2, we conclude $E_a\le E_b$.

%%%%%%%%%%%%%%%%%%%%%%%%%%%%%%%%%%%%%%%%%%%%%%%%%%%%%%%%%%%%%%%%%%%%%%%%%%%%%%%%%%%%%%%%
%%%%%%%%%%%%%%%%%%%%%%%%%%%%%%%%%%%%%%%%%%%%%%%%%%%%%%%%%%%%%%%%%%%%%%%%%%%%%%%%%%%%%%%%
\subsubsection*{An example}
%%%%%%%%%%%%%%%%%%%%%%%%%%%%%%%%%%%%%%%%%%%%%%%%%%%%%%%%%%%%%%%%%%%%%%%%%%%%%%%%%%%%%%%%
%%%%%%%%%%%%%%%%%%%%%%%%%%%%%%%%%%%%%%%%%%%%%%%%%%%%%%%%%%%%%%%%%%%%%%%%%%%%%%%%%%%%%%%%
To demonstrate Theorem 1 we choose the laser--dressed potential $V_a$ \cite{laser1, laser2, Hall3} and the exponential potential $V_b$  in the form
\begin{equation*}
V_a=-\frac{\alpha}{(x^2+a^2)^{1/2}} \qquad \text{and} \qquad 
V_b=-\beta e^{-b|x|},
\end{equation*}
with $\alpha=0.61362$, $a=0.62$, $\beta=0.8$, and $b=0.41$; see Figure~2.
\begin{figure}
\centering{\includegraphics[height=13cm,width=8cm,angle=-90]{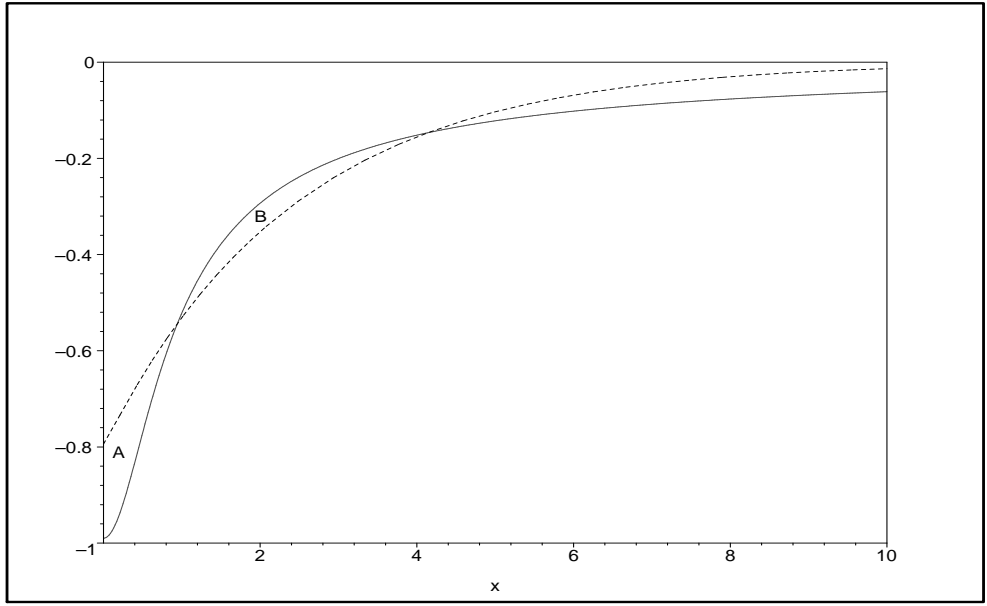}}
\caption{The laser--dressed potential $V_a$ full line and the exponential potential $V_b$ dotted line.}
\end{figure}
Thus $V_a(0)=-0.989$ and $V_b(0)=-0.8$, and taking $m=1$, the condition $-V_0\le 2m$ is satisfied. The graphs of $V_a$ and $V_b$ intersect at $x_1=0.94437$ and $x_2=4.13782$. Then we calculate areas $A$ and $B$
\begin{equation*}
A=\int_0^{x_1} (V_b - V_a)dx=0.11456
\end{equation*}
and 
\begin{equation*}
B=\int_{x_1}^{x_2} (V_a - V_b)dx=0.11455.
\end{equation*}
Since $A>B$ we have $g>0$ therefore according to Corollary 1 we should have $E_a\le E_b$, which we verify by calculating accurate numerical eigenvalues, i.e. $E_a=0.45657\le E_b=0.52332$.

%%%%%%%%%%%%%%%%%%%%%%%%%%%%%%%%%%%%%%%%%%%%%%%%%%%%%%%%%%%%%%%%%%%%%%%%%%%%%%%%%%%%%%%%
%%%%%%%%%%%%%%%%%%%%%%%%%%%%%%%%%%%%%%%%%%%%%%%%%%%%%%%%%%%%%%%%%%%%%%%%%%%%%%%%%%%%%%%%
\section{Dirac equation in $d>1$ dimensions}
%%%%%%%%%%%%%%%%%%%%%%%%%%%%%%%%%%%%%%%%%%%%%%%%%%%%%%%%%%%%%%%%%%%%%%%%%%%%%%%%%%%%%%%%
%%%%%%%%%%%%%%%%%%%%%%%%%%%%%%%%%%%%%%%%%%%%%%%%%%%%%%%%%%%%%%%%%%%%%%%%%%%%%%%%%%%%%%%%
For a central potential in $d>1$ dimensions the Dirac equation can be written \cite{Bjorken} in natural units $\hbar=c=1$ as
\begin{equation*}
i\frac{\partial \Psi}{\partial t} =H\Psi,\quad {\rm where}\quad  H=\sum_{s=1}^{d}{\alpha_{s}p_{s}} + m\beta+V,
\end{equation*}
where $m$ is the mass of the particle, $V$ is an attractive spherically--symmetric potential, which will be defined later, and $\{\alpha_{s}\}$ and $\beta$  are Dirac matrices, which satisfy anti--commutation relations; the identity matrix is implied after the potential $V$. For stationary states, some algebraic calculations in a suitable basis, the details of which may be found in Refs. \cite{Gu, Dong, jiang, salazar, yasuk}, lead to a pair of first--order linear differential equations in two radial wave functions $\{\psi_1, \psi_2\}$, namely
\begin{subnumcases}{}
\label{dcel1}
\psi_1'=(m+E-V)\psi_2-\frac{k_d}{r}\psi_1,\\
\label{dcer1}
\psi_2'=(m-E+V)\psi_1+\frac{k_d}{r}\psi_2,
\end{subnumcases}
where $r = \|\mb{r}\|$, prime $'$ denotes the derivative with respect to $r$, $k_d=\tau\left(j+\frac{d-2}{2}\right)$, $\tau = \pm 1$, and $j=1/2,\ 3/2,\ 5/2,\ \ldots$.  We note that the variable $\tau$ is sometimes written $\omega$, as, for example in the book by Messiah \cite{messiah}, and the radial functions are often written $\psi_1 = G$ and $\psi_2 = F,$ as in the book by Greiner \cite{greiner}. For $d > 1,$ these functions vanish at $r = 0$, and, for bound states, they may be normalized by the relation 
\begin{equation*}
(\psi_1,\psi_1) + (\psi_2,\psi_2) = \int\limits_0^{\infty}(\psi_1^2 + \psi_2^2)dr = 1.
\end{equation*}
We use inner products {\it without} the radial measure $r^{(d-1)}$ because the factor $r^{\frac{(d-1)}{2}}$ is already built in to each radial function.  We shall assume that the potential $V$ is such that there is a discrete energy eigenvalue $E$ and that equations (\ref{dcel1})--(\ref{dcer1}) are the eigenequations for the corresponding radial eigenstates. Throughout this paper we will consider only potentials which vanish at infinity, thus the above system at infinity becomes
\begin{subnumcases}{}
\label{dcel15}
\psi_1'=(m+E)\psi_2,\\
\label{dcer15}
\psi_2'=(m-E)\psi_1.
\end{subnumcases}
Let us assume that $\psi_1\ge 0$ before vainshing, then it follows from (\ref{dcel15}) that $(m+E)\psi_2\le 0$. Thus either $m+E>0$ and $\psi_2\le 0$ or $m+E<0$ and $\psi_2\ge 0$. By considering equation (\ref{dcer15}), the first case leads to $\psi_2'\ge 0$ and $m-E>0$, so $-m<E<m$. The second case leads to $\psi_2'\le 0$ and $m-E<0$; but this is the contradiction: it follows from $m+E<0$ that $E<0$ and from $m-E<0$ that $E>0$. Therefore we conclude that if the potential $V$ vanishes at infinity then the discrete energy $E$ is such that $-m<E<m$.

%%%%%%%%%%%%%%%%%%%%%%%%%%%%%%%%%%%%%%%%%%%%%%%%%%%%%%%%%%%%%%%%%%%%%%%%%%%%%%%%%%%%%%%%
%%%%%%%%%%%%%%%%%%%%%%%%%%%%%%%%%%%%%%%%%%%%%%%%%%%%%%%%%%%%%%%%%%%%%%%%%%%%%%%%%%%%%%%%
\section{Refined comparison theorems for the Dirac equation in $d> 1$ dimensions}
%%%%%%%%%%%%%%%%%%%%%%%%%%%%%%%%%%%%%%%%%%%%%%%%%%%%%%%%%%%%%%%%%%%%%%%%%%%%%%%%%%%%%%%%
%%%%%%%%%%%%%%%%%%%%%%%%%%%%%%%%%%%%%%%%%%%%%%%%%%%%%%%%%%%%%%%%%%%%%%%%%%%%%%%%%%%%%%%%
As in one--dimensional case we need to know some characteristics of the nodeless state of the Dirac coupled equations (\ref{dcel1})--(\ref{dcer1}). It follows from the Nodal Theorem of Ref. \cite{nod} that in the state with no nodes $k_d<0$, and either $\psi_1\ge 0$ and $\psi_2\le 0$ or $\psi_1\le 0$ and $\psi_2\ge 0$ for $r\in[0,\ \infty)$; so from now on without loss of generality we suppose $k_d<0$ and  $\psi_1\ge 0$ and $\psi_2\le 0$ for $r\in[0,\ \infty)$.  In Figure 3 we present an illustration of a node free state.
\begin{figure}
\centering{\includegraphics[height=13cm,width=5cm,angle=-90]{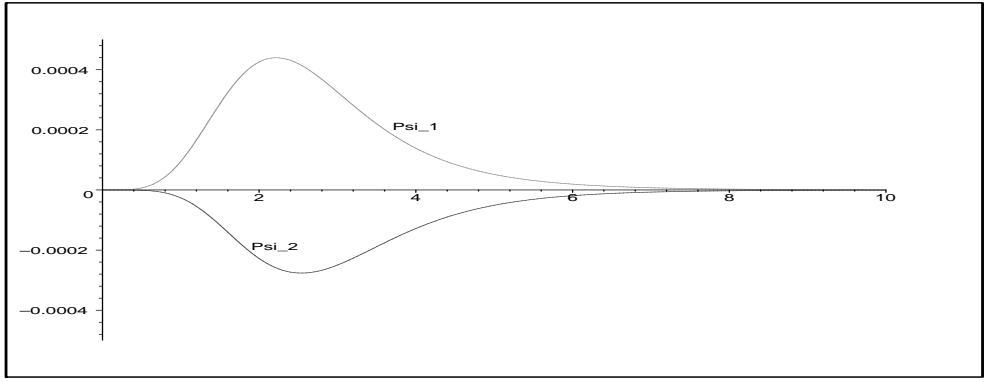}}
\caption{The Dirac radial wave functions $\psi_1$ and $\psi_2$ at the bottom of an angular--momentum subspace labelled by $j$, corresponding to the Woods--Saxon potential \cite{WS} in the form $V=-\cfrac{v}{1+e^{\frac{r-R}{a}}}$, with $v=4$, $R=2$, $a=1.2$, $\tau=-1$, $d=8$, $j=3/2$, $m=1$, and the energy eigenvalue is $E=0.62317$.}
\end{figure}
Using (\ref{dcel1})--(\ref{dcer1}) and following the same argument as in one--dimensional case, we can obtain the corresponding relation for two comparison potentials $V_a$ and $V_b$
\begin{equation}\label{expr5}
(E_b - E_a)\int_0^\infty (\psi_{1a}\psi_{1b} + \psi_{2a}\psi_{2b})dr=\int_0^\infty(V_b-V_a) (\psi_{1a}\psi_{1b} + \psi_{2a}\psi_{2b})dr.
\end{equation}
From equation (\ref{expr5}) we can eventually recover the basic comparison theorem \cite{p75} to the effect that if the radial components of the Dirac spinor are node free and $V_a\le V_b$, then $E_a\le E_b$.

%%%%%%%%%%%%%%%%%%%%%%%%%%%%%%%%%%%%%%%%%%%%%%%%%%%%%%%%%%%%%%%%%%%%%%%%%%%%%%%%%%%%%%%%
\subsection{Bounded Potentials}
%%%%%%%%%%%%%%%%%%%%%%%%%%%%%%%%%%%%%%%%%%%%%%%%%%%%%%%%%%%%%%%%%%%%%%%%%%%%%%%%%%%%%%%%
Here we suppose that the potential $V$ in the Dirac equations (\ref{dcel1})--(\ref{dcer1}) is such that
\begin{eqnarray*}
&(\rmn{1})& V\ \text{is nonpositive and bounded, i.e.} \ V_0\le V\le 0,\ \text{where}\ V_0=V(0);\\
\vspace{9mm}
&(\rmn{2})& V\ \text{vanishes at infinity, thus}\ \lim_{r\to\infty}V=0;\\
\vspace{9mm}
&(\rmn{3})& V\ \text{is attractive, that is monotone nondecreasing on} \  [0, \infty);\ \text{thus, when $V$ is differentiable,}\  V'\ge 0.
\vspace{9mm}
\end{eqnarray*}

We shall first prove the following lemma, which characterizes the behaviour of the Dirac radial wave functions at the bottom of the angular--momentum subspace labelled by $j$: for these nodefree states \cite{nod}, $k_d=-\left(j+\cfrac{d-2}{2}\right)$, where $j=1/2,\ 3/2,\ 5/2,\ \ldots$ and $d=2,\ 3,\ 4,\ \ldots$.   

\medskip

\noindent{\bf Lemma 1:} ~~{\it the Dirac radial spinor components $\psi_1$ and $\psi_2$ at the bottom of an angular--momentum subspace labelled by $j$, which satisfy (\ref{dcel1})--(\ref{dcer1}), for the bounded potential $V$ are such that
\begin{equation*}
\left(\frac{\psi_1}{r^{|k_d|}}\right)'\le 0 \quad \text{and} \quad
\left(\frac{\psi_2}{r^{|k_d|+1}}\right)'\ge 0, \quad r\in[0, \ \infty).
\end{equation*}} 

\medskip

\noindent{\bf Proof:} Near the origin the system (\ref{dcel1})--(\ref{dcer1}) may be rewritten as: 
\begin{subnumcases}{}
\psi_1''=\psi_1\left(\frac{k_d(k_d+1)}{r^2}+m^2-(E-V_0)^2\right),\\
\psi_2''=\psi_2\left(\frac{k_d(k_d-1)}{r^2}+m^2-(E-V_0)^2\right).
\end{subnumcases}
Solutions of these equations involve Bessel functions. Hence for small $r$ we can approximate them by simple powers in the following form
\begin{subnumcases}{}
\label{powl}
\psi_1=c_1r^{q_1},\\
\label{powr}
\psi_2=c_2r^{q_2},
\end{subnumcases}
where $c_1$ and $c_2$ are constants of integration and parameters $q_1$ and $q_2$ are positive since both wave functions must vanish at the origin. After substituting (\ref{powl})--(\ref{powr}) into (\ref{dcel1})--(\ref{dcer1}) and dividing one equation by the other we obtain the following relation
\begin{equation*}
(q_1+k_d)(q_2-k_d)=(m+E-V_0)(m-E+V_0)r^2,
\end{equation*}
which in the limit as $r$ approaches $0$ reduces to
\begin{equation*}
(q_1+k_d)(q_2-k_d)=0.
\end{equation*}
Since $k_d<0$ and $q_1>0$, it follows from the above expression that $q_1=-k_d$. Then equation (\ref{dcer1}) becomes:
\begin{equation*}
c_2(q_2-k_d)r^{q_2-1}=c_1(m-E+V_0)r^{-k_d}.
\end{equation*}
Equating the powers of $r$ we obtain $q_2=1-k_d$. Also one finds 
\begin{equation*}
\frac{c_1}{c_2}=\frac{1-2k_d}{m-E+V_0}.
\end{equation*}
According to $(ii)$ and $(iii)$, $\lim\limits_{r\to\infty}(m-E+V)=m-E>0$, meanwhile $m-E+V_0<0$, thus the quantity $m-E+V$ changes sign exactly once \cite{nod} for $r>0$. The ratio $c_1/c_2<0$, which means that $\psi_1$ and $\psi_2$ have opposite signs (this is in agreement with our assumption for the nodeless state, that is $\psi_1\ge 0$ and $\psi_2\le 0$ on $[0,\ \infty)$). Finally, we conclude that near the origin radial wave functions behave as
\begin{subnumcases}{}
\label{beh1}
\psi_1=C_1r^{-k_d},\\
\label{beh2}
\psi_2=C_2r^{-k_d+1}.
\end{subnumcases}

Now let us make the following substitution $\psi_1=r^{-k_d}R_1$ and $\psi_2=r^{-k_d+1}R_2$, then the system of equations (\ref{dcel1})--(\ref{dcer1}) becomes
\begin{subnumcases}{}
\label{dcel2}
R_1'=(m+E-V)rR_2,\\
\label{dcer2}
R_2'=(m-E+V)\frac{R_1}{r}+\frac{2k_d-1}{r}R_2.
\end{subnumcases}
According to $(i)$ and (\ref{dcel2}), $R_1'\le 0$ which is equivalent to the lemma's first inequality. Since $m-E+V$ has to change sign from negative to positive, thus for large $r$, according to (\ref{dcer2}), $R_2'\ge 0$. In order to determine the behaviour of the $R_2'$ near the origin, we expand $R_1$ and $R_2$ in power series, i. e. $R_1=a_0+a_1r+a_2r^2+a_3r^3+\ldots$ and $R_2=b_0+b_1r+b_2r^2+b_3r^3+\ldots$, then system (\ref{dcel2})--(\ref{dcer2}) implies
\begin{subnumcases}{}
\label{dcel3}
R_1=a_0+a_2r^2+\ldots,\\
\label{dcer3}
R_2=b_0+b_2r^2+\ldots,
\end{subnumcases}
where if $a_0>0$, then $b_0<0$, $a_2<0$, and $b_2>0$. Thus $R_2'\ge 0$ near zero. 

Let us suppose that the function $R_2$ is decreasing on some interval $(r_1,\ r_2)$, i. e. $R_2'<0$ on $(r_1,\ r_2)$ and $R_2'(r_1)=R_2'(r_2)=0$. Then it follows from the Rolle's Theorem that there is at least one number $r_c\in(r_1,\ r_2)$ such that $R_2''(r_c)=0$, which is equivalent to
\begin{equation*}
\left((m-E+V)\frac{R_1}{r}+\frac{2k_d-1}{r}R_2\right)'=0 
\quad\text{at}\quad r=r_c
\end{equation*}
or, using (\ref{dcer2}),
\begin{equation*}
V'R_1 + (m-E+V)R_1'+(2k_d-2)R_2'=0
\quad\text{at}\quad r=r_c.
\end{equation*}
In the above expression first two terms are nonnegative and $(2k_d-2)R_2'$ is strictly positive, which yields a contradiction. Hence $R_2'\ge 0$ on $r\in[0, \ \infty)$ and this corresponds to the lemma's second inequality.   

\hfill $\Box$

\medskip

\noindent{\bf Theorem 3:} ~~{\it The potential $V$, satisfies $(\rmn{1})$--$(\rmn{2})$, $V_0\ge-2m$, and has $r^{2|k_d|}$-weighted area, if 
\begin{equation}\label{th5}
\eta(r)=\int_0^r (V_b(t)-V_a(t))t^{2|k_d|}dt\ge 0, \quad r\in [0,\ \infty),
\end{equation}
then we have $E_a\le E_b$.} 

\medskip

\noindent{\bf Proof:} Let us integrate the right side of (\ref{expr5}) by parts in the following way 
\begin{equation*}
\int_0^\infty(V_b-V_a) (\psi_{1a}\psi_{1b} + \psi_{2a}\psi_{2b})dr=
\left.(\psi_{1a}\psi_{1b} + \psi_{2a}\psi_{2b})\frac{\eta}{r^{2|k_d|}}\right|_0^\infty-
\int_0^\infty\eta\left(\frac{\psi_{1a}\psi_{1b} + \psi_{2a}\psi_{2b}}{r^{2|k_d|}}\right)'dr,
\end{equation*}
where $\eta(r)$ is defined by (\ref{th5}). Since $\eta(0)=0$ and $\lim\limits_{r\to\infty}\psi_{1}=\lim\limits_{r\to\infty}\psi_{2}=0$ with the respective asymptotic forms (\ref{beh1})--(\ref{beh2}), relation (\ref{expr5}) becomes
\begin{equation}\label{eq34}
(E_b - E_a)\int_0^\infty (\psi_{1a}\psi_{1b} + \psi_{2a}\psi_{2b})dr=
-\int_0^\infty\eta\left(\frac{\psi_{1a}\psi_{1b} +\psi_{2a}\psi_{2b}}{r^{2|k_d|}}\right)'dr.
\end{equation}
Using (\ref{dcel1})--(\ref{dcer1}) we find
\begin{equation*}
\left(\frac{\psi_{1a}\psi_{1b} +\psi_{2a}\psi_{2b}}{r^{2|k_d|}}\right)'=
W_1\frac{\psi_{1b}\psi_{2a}}{r^{2|k_d|}} + W_2\frac{\psi_{1a}\psi_{2b}}{r^{2|k_d|}}+
4k_d\frac{\psi_{2a}\psi_{2b}}{r^{2|k_d|+1}},
\end{equation*}
where
\begin{equation*}
W_1=2m+E_a-E_b-V_a+V_b \quad \text{and} \quad 
W_2=2m-E_a+E_b+V_a-V_b.
\end{equation*}
We know that quantity $m-E+V$ changes sign from negative to positive. When $m-E+V\le 0$ we have $W_1\ge 3m-E_b+V_b$. Since $-m<E<m$ and $V\ge -2m$, then $W_1\ge 3m-E_b+V_b> 0$. When $m-E+V> 0$ it is straightforward that $W_1\ge m+E_a-V_a>0$. Therefore $W_1>0$ for $r\in[0,\ \infty)$. Similarly it can be shown that $W_2>0$. Thus the derivative $\left(\cfrac{\psi_{1a}\psi_{1b} +\psi_{2a}\psi_{2b}}{r^{2|k_d|}}\right)'\le 0$ and according to the theorem's assumption (\ref{th5}) and equation (\ref{eq34}) we have $E_a\le E_b$.

\hfill $\Box$ 

\medskip

We note that above theorem, as well as Theorem 1, does not require a nondecreasing potential $V$ on $[0, \infty)$, i. e. $V$ can decrease on some intervals. As in the one--dimensional case, if we know more precise behaviour of the comparison potentials, we can state simpler sufficient conditions:

\medskip

\noindent{\bf Corollary 3:} ~~{\it If the potentials cross over once, say at $r_1$, and $V_a \le V_b$ for $r\in [0,\ r_1]$, and
\begin{equation*}
\eta(\infty)=\int_0^\infty (V_b-V_a)r^{2|k_d|} dr\ge 0, 
\end{equation*}
then $E_a\le E_b$. If the potentials cross over twice, say at $r_1$ and $r_2$, $r_1<r_2$, $V_a\le V_b$ for $r\in [0,\ r_1]$, and
\begin{equation*}
\eta(r_2)=\int_0^{r_2} (V_b-V_a)r^{2|k_d|} dr\ge 0,   
\end{equation*}
then $E_a\le E_b$.} 

\medskip

We can extend the above corollary in the following way: assume that comparison potentials have $n$ intersections, $n=1,\ 2,\ 3,\ \ldots$, and $V_a\le V_b$ on $r\in[0,\ r_1]$. Also assume that $\int_{r_i}^{r_{i+1}}|V_b-V_a|r^{2|k_d|}dr$, $i=1,\ 2,\ 3,\ \ldots,\ n$ is nonincreasing sequence (if $n$ is odd then $\int_{r_{n-1}}^{r_n}|V_b-V_a|r^{2|k_d|}dr\ge\int_{r_n}^\infty|V_b-V_a|r^{2|k_d|}dr$ ), hence $\int_0^r (V_b(t)-V_a(t))t^{2|k_d|}dt\ge 0$, $r\in[0,\ \infty)$, and we conclude $E_a\le E_b$.

\medskip

\noindent{\bf Theorem 4:} ~~{\it The potential $V$ satisfies $(\rmn{1})$--$(\rmn{3})$, and has $\psi_{1i}r^{|k_d|}$ and $\psi_{2i}r^{|k_d|+1}$--weighted areas, if 
\begin{equation}\label{th6}
\lambda_1(r)=\int_0^r (V_b(t)-V_a(t))\psi_{1i}(t)t^{|k_d|}dt\ge 0 \quad \text{and} \quad
\lambda_2(r)=\int_0^r (V_b(t)-V_a(t))(-\psi_{2i}(t))t^{|k_d|+1}dt\ge 0, 
\end{equation}
$r\in [0,\ \infty),$ where $i$ is either $a$ or $b$, then we have $E_a\le E_b$.} 

\medskip

\noindent{\bf Proof:} We prove the theorem for $i=b$; for $i=a$, the proof is the same. After integrating the right side of (\ref{expr5}) by parts we obtain 
\begin{equation*}
\int_0^\infty(V_b-V_a) (\psi_{1a}\psi_{1b} + \psi_{2a}\psi_{2b})dr=
\left[\lambda_1\frac{\psi_{1a}}{r^{|k_d|}}+\lambda_2\frac{\psi_{2a}}
{r^{|k_d|+1}}\right]_0^\infty
-\int_0^\infty \left[\lambda_1\left(\frac{\psi_{1a}}{r^{|k_d|}}\right)' + \lambda_2\left(\frac{\psi_{2a}}{r^{|k_d|+1}}\right)'\right]dr,
\end{equation*}
where $\lambda_1(r)$ and $\lambda_2(r)$ are defined by (\ref{th6}) for $i=b$. Since $\lambda_1(0)=\lambda_2(0)=0$ and $\lim\limits_{r\to\infty}\psi_{1a}=\lim\limits_{r\to\infty}\psi_{2a}=0$ with the respective asymptotic forms (\ref{beh1})--(\ref{beh2}), expression (\ref{expr5}) becomes
\begin{equation*}
(E_b - E_a)\int_0^\infty (\psi_{1a}\psi_{1b} + \psi_{2a}\psi_{2b})dr=
-\int_0^\infty \left[\lambda_1\left(\frac{\psi_{1a}}{r^{|k_d|}}\right)' + \lambda_2\left(\frac{\psi_{2a}}{r^{|k_d|+1}}\right)'\right]dr.
\end{equation*}
Then Lemma 1 and the theorem's assumptions ensure that $E_a\le E_b$.

\hfill $\Box$

\medskip

\noindent{\bf Corollary 4:} ~~{\it If the potentials cross over once, say at $r_1$, $V_a\le V_b$ for $r\in [0,\ r_1]$, and
\begin{equation*}\label{concl5}
\lambda_1(\infty)=\int_0^\infty (V_b-V_a)\psi_{1i}r^{|k_d|}dr\ge 0 \quad \text{and} \quad
\lambda_2(\infty)=\int_0^\infty (V_b-V_a)(-\psi_{2i})r^{|k_d|+1}dr\ge 0, 
\quad i=a\ or\ b,
\end{equation*}
then $E_a\le E_b$. If the potentials cross over twice, say at $r_1$ and $r_2$, $r_1<r_2$, $V_a\le V_b$ for $r\in [0,\ r_1]$, and
\begin{equation*}\label{concl6}
\lambda_1(r_2)=\int_0^{r_2} (V_b-V_a)\psi_{1i}r^{|k_d|}dr\ge 0 \quad \text{and} \quad
\lambda_2(r_2)=\int_0^{r_2} (V_b-V_a)(-\psi_{2i})r^{|k_d|+1}dr\ge 0, 
\quad i=a\ or\ b,  
\end{equation*}
then $E_a\le E_b$.} 

\medskip

As before we can generalize Corollary 4 to allow $n$ intersections, i.e. if $V_a\le V_b$ on $r\in[0,\ r_1]$ and both sequences of absolute areas $\int_{r_i}^{r_{i+1}}|(V_b-V_a)\psi_{1i}r^{|k_d|}|dr$ and $\int_{r_i}^{r_{i+1}}|(V_b-V_a)\psi_{2i}r^{|k_d|+1}|dr$, $i=a$ or $b$, are nonincreasing (if $n$ is odd then we assume $\int_{r_{n-1}}^{r_n}|(V_b-V_a)\psi_{1i}r^{|k_d|}|dr\ge\int_{r_n}^\infty|(V_b-V_a)
\psi_{1i}r^{|k_d|}|dr$ and $\int_{r_{n-1}}^{r_n}|(V_b-V_a)\psi_{2i}r^{|k_d|+1}|dr\ge\int_{r_n}^\infty|(V_b-V_a)
\psi_{2i}r^{|k_d|+1}|dr$), then integrals $\int_0^r (V_b(t)-V_a(t))\psi_{1i}(t)t^{|k_d|}dt\ge 0$ and $\int_0^r (V_b(t)-V_a(t))(-\psi_{2i}(t))t^{|k_d|+1}dt\ge 0$ for $r\in[0,\ \infty)$, thus $E_a\le E_b$.

%%%%%%%%%%%%%%%%%%%%%%%%%%%%%%%%%%%%%%%%%%%%%%%%%%%%%%%%%%%%%%%%%%%%%%%%%%%%%%%%%%%%%%%%
\subsubsection*{An example}
%%%%%%%%%%%%%%%%%%%%%%%%%%%%%%%%%%%%%%%%%%%%%%%%%%%%%%%%%%%%%%%%%%%%%%%%%%%%%%%%%%%%%%%%
As an example we consider Theorem 3, in particular Corollary 3 for the case of many intersections. We take the following comparison potentials
\begin{equation*}
V_a=-\frac{\alpha}{ur^3+a}\left(1+\frac{v\sin(\kappa r^3 + s)}{\kappa r^3 + s}\right) \qquad \text{and} \qquad 
V_b=-\frac{\beta}{wr^3+b}.
\end{equation*}
If $\alpha=\beta$, $a=s=b$, $u=\kappa=w$, $\tau=-1$, $j=1/2$, and $d=3$ the substitution $z=ur^3+a$ transforms the integral (\ref{th5}) into
\begin{equation*}
\int_0^\infty (V_b-V_a)r^{2}dt=\frac{\alpha v}{3u}\int_a^\infty\frac{\sin z}{z^2}dz. 
\end{equation*}
The integrand $I=\cfrac{\sin z}{z^2}$ is plotted on Figure 4. 
\begin{figure}
\centering{\includegraphics[height=15cm,width=6cm,angle=-90]{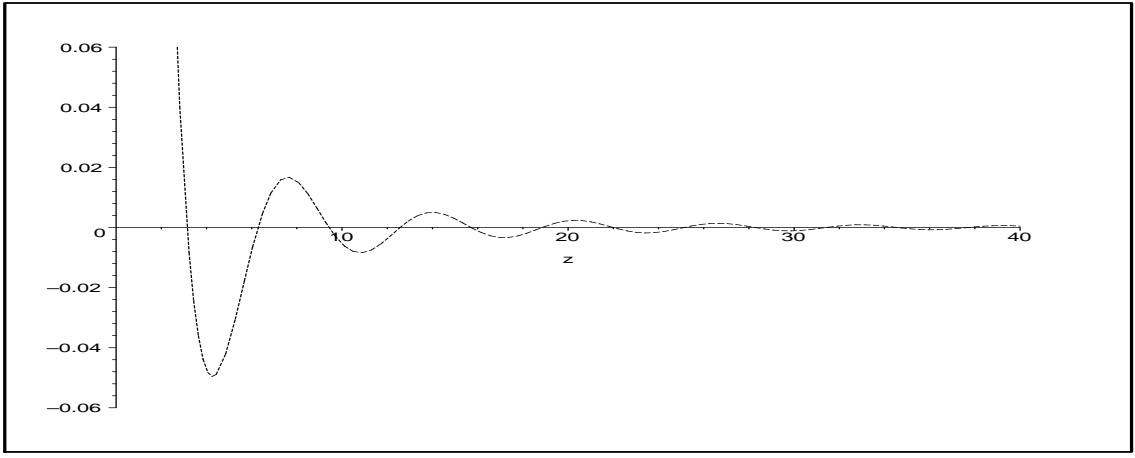}}
\caption{The graph of the integrand $I=\cfrac{\sin z}{z^2}$.}
\end{figure}
Choosing $a=2.04$, and calculating numerical values, we find that the first area is bigger then the second one:
\begin{equation*}
\int_a^\pi \frac{|\sin z|}{z^2}dz =0.0965>\int_\pi^{2\pi} \frac{|\sin z|}{z^2}dz=0.0962.
\end{equation*} 
The $\sin z$ is the periodic function, thus $|\sin x|=|\sin y|$ where $x\in[(k-1)\pi,\ k\pi]$ and $y=x+\pi$, $k=3,\ 4,\ 5,\ \ldots$, then it is clear that 
\begin{equation*}
\int_{(k-1)\pi}^{k\pi} \frac{|\sin z|}{z^2}dz >\int_{k\pi}^{(k+1)\pi} \frac{|\sin z|}{z^2}dz.
\end{equation*}
Therefore 
\begin{equation*}
\int_0^\infty (V_b-V_a)r^{2}dt\ge 0,
\end{equation*}
because successive positive and negative areas of the integrand do not increase in absolute value. Thus $\eta>0$ and by Theorem 3 we have $E_a\le E_b$.  
This prediction is verified by accurate numerical calculations:
for $\alpha=3.4$, $a=2.04$, $v=0.4$, and $u=7$ the comparison potentials intersect at infinitely many points (see Figure 5), with $m=1$ so that condition $-2m\le V_0$ is satisfied.
\begin{figure}[ht]
\begin{center}$
\begin{array}{cc}
\includegraphics[width=2.5in, angle=-90]{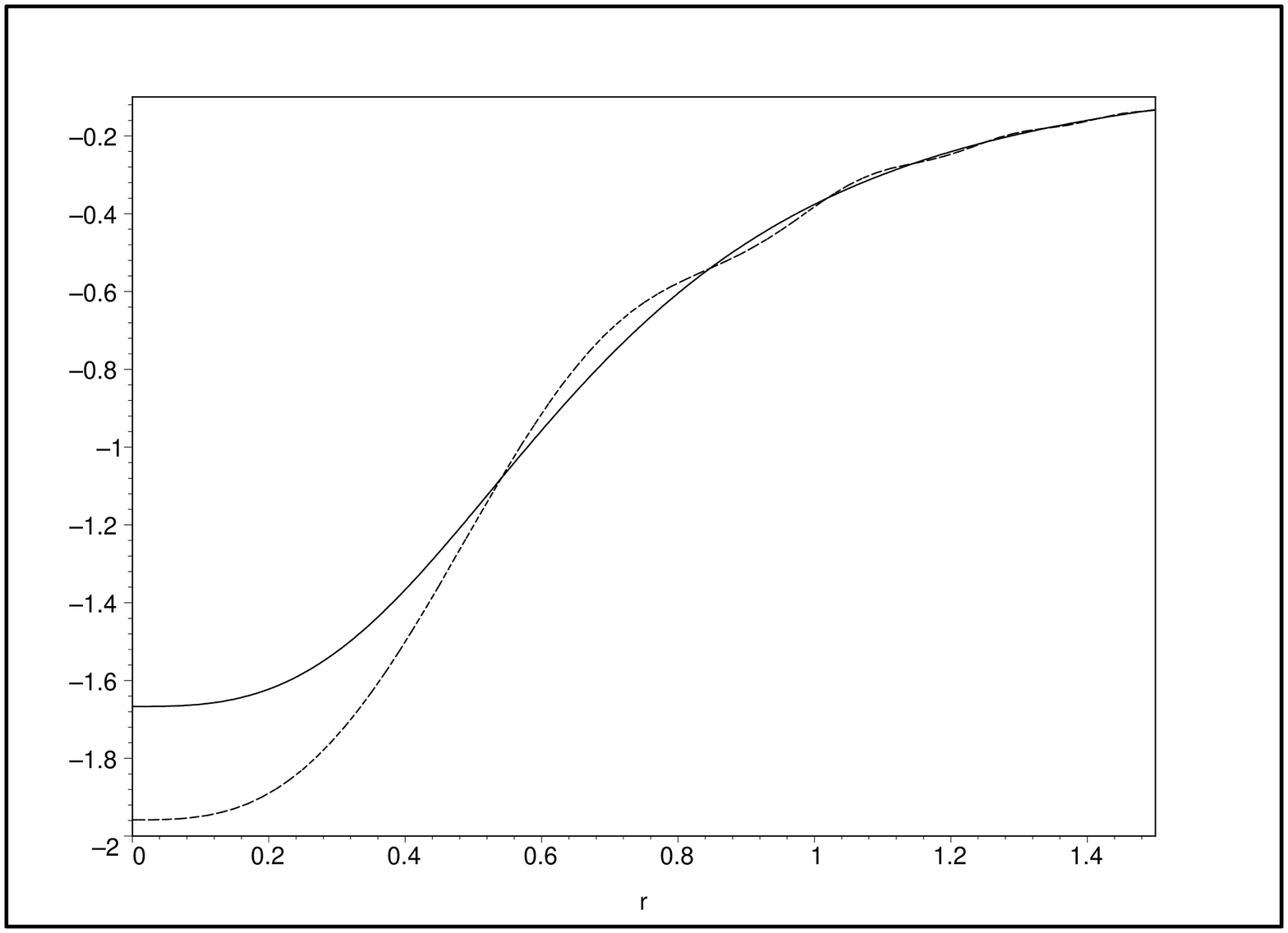} & 
\includegraphics[width=2.5in, angle=-90]{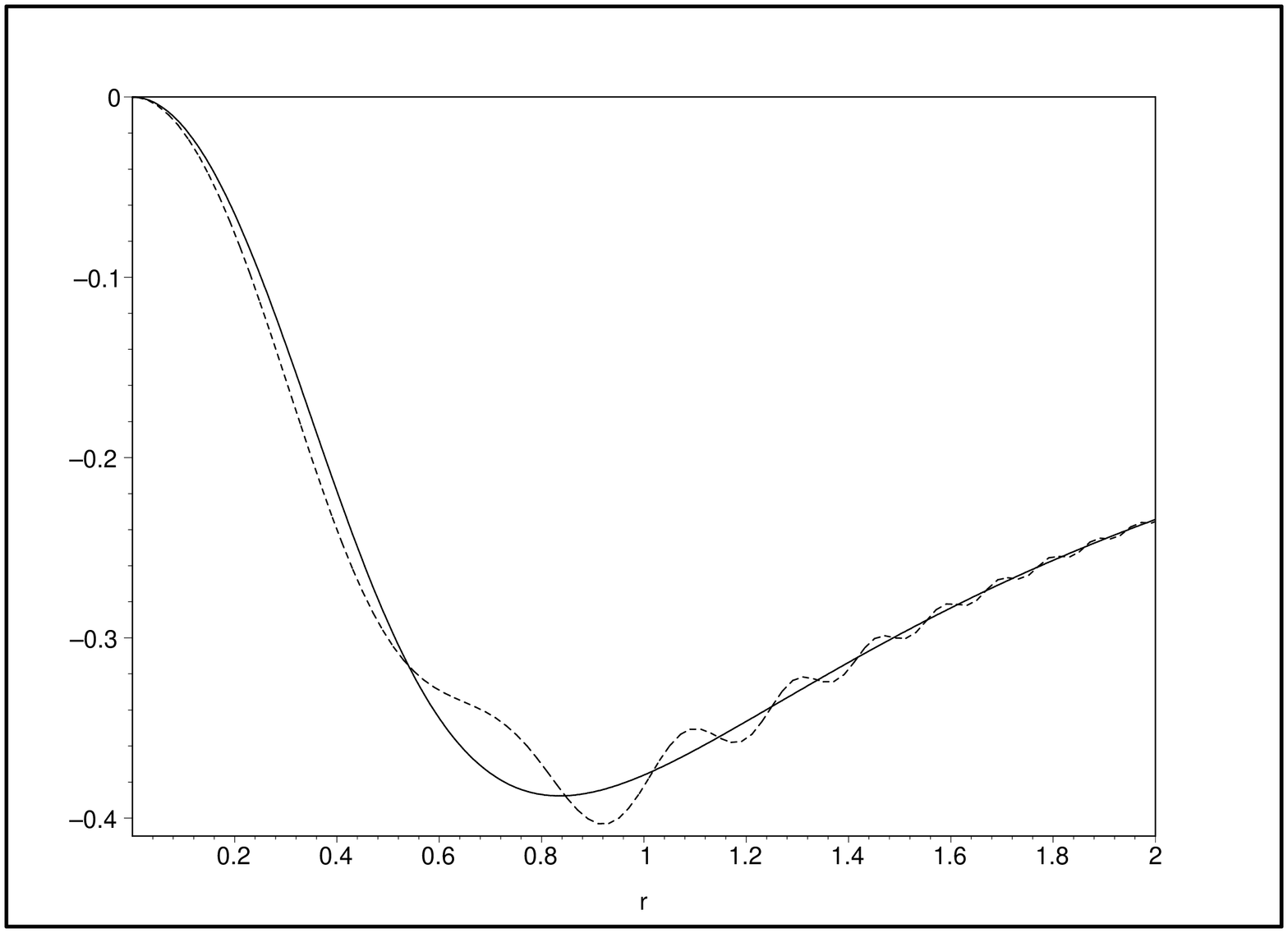}
\end{array}$
\end{center}
\caption{Left graph: potential $V_a$ dashed line and $V_b$ full line. Right graph: function $V_ar^{2|k_d|}$ dashed line and $V_br^{2|k_d|}$ full line.}
\end{figure}
Accurate numerical eigenvalues are $E_a=0.99427\le E_b=0.99542$.

%%%%%%%%%%%%%%%%%%%%%%%%%%%%%%%%%%%%%%%%%%%%%%%%%%%%%%%%%%%%%%%%%%%%%%%%%%%%%%%%%%%%%%%%
\subsection{Unbounded Potentials}
%%%%%%%%%%%%%%%%%%%%%%%%%%%%%%%%%%%%%%%%%%%%%%%%%%%%%%%%%%%%%%%%%%%%%%%%%%%%%%%%%%%%%%%%
We consider a class of unbounded potentials of the form $V(r)=-\cfrac{f(r)}{r}$, where the bounded factor $f$ satisfies:
\begin{eqnarray*}
&(\rmn{4})& f\ \text{is nonnegative and bounded, i.e.} \ f_0\ge f\ge 0,\ \text{where}\ f_0=\lim\limits_{r\to 0^+}f;\\
\vspace{9mm}
&(\rmn{5})& f\ \text{is monotone nonincreasing on} \  [0, \infty),\ \text{so}\ f'\le 0;\\
\vspace{9mm}
&(\rmn{6})& V\ \text{vanishes at infinity, so}\ \lim_{r\to\infty}\cfrac{f}{r}=0.
\end{eqnarray*}
For instance, for the Coulomb potential $V=-\cfrac{v}{r}$, the function $f=v$, for the Yukawa potential \cite{Yuk} $V=-\cfrac{v}{re^{\lambda r}}$, the function $f=\cfrac{v}{e^{\lambda r}}$, for the Hulth\'en potential \cite{Hult} $V=-\cfrac{v}{e^{\lambda r}-1}$, the function $f=\cfrac{vr}{e^{\lambda r}-1}$, and so on.

\medskip

\noindent{\bf Lemma 2:} ~~{\it the Dirac radial spinor components $\psi_1$ and $\psi_2$ at the bottom of an angular--momentum subspace labelled by $j$, which satisfy (\ref{dcel1})--(\ref{dcer1}), for the potential $V=-\cfrac{f}{r}$ are such that
\begin{equation*}
\left(\frac{\psi_1}{r^{|k_d|}}\right)'\le 0 \quad \text{and} \quad
\left(\frac{\psi_2}{r^{|k_d|}}\right)'\ge 0, \quad r\in[0, \ \infty).
\end{equation*}} 

\medskip

\noindent{\bf Proof:} For small $r$ analysis of the Dirac coupled equations (\ref{dcel1})--(\ref{dcer1}) yields the asymptotic forms:
\begin{subnumcases}{}
\psi_1''=\psi_1\frac{k^2_d-f^2_0}{r^2}-\frac{\psi '_1}{r},\\
\psi_2''=\psi_2\frac{k^2_d-f_0^2}{r^2}-\frac{\psi '_2}{r}.
\end{subnumcases}
These are Cauchy–-Euler equations with solution in the form \cite{C-E}:
\begin{subnumcases}{}
\label{powl1}
\psi_1=c_1r^{\gamma},\\
\label{powr1}
\psi_2=c_2r^{\gamma},
\end{subnumcases}
where $c_1$ and $c_2$ are constants of integration, and the parameter $\gamma$ has to be positive because the wave functions must vanish at the origin. Substitution of (\ref{powl1})--(\ref{powr1}) into (\ref{dcel1})--(\ref{dcer1}) yields
\begin{subnumcases}{}
\nonumber
\gamma=f_0c_2/c_1-k_d,\\
\nonumber
\gamma=-f_0c_1/c_2+k_d.
\end{subnumcases}
The solution of the above system is: $\gamma=\sqrt{k_d^2-f_0^2}$ and $\cfrac{c_2}{c_1}=\cfrac{k_d\pm\sqrt{k_d^2-f_0^2}}{f_0}$. As in sec. A, $c_2/c_1<0$, which is in agreement with our assumption for the nodeless state. Therefore near $0$ the wave functions behave as
\begin{subnumcases}{}
\nonumber
\psi_1=c_1r^{\sqrt{k_d^2-f_0^2}},\\
\nonumber
\psi_2=c_2r^{\sqrt{k_d^2-f_0^2}}.
\end{subnumcases}
We now substitute $\psi_1=r^{-k_d}P_1$ and $\psi_2=r^{-k_d}P_2$, into (\ref{dcel1})--(\ref{dcer1}) to obtain:  
\begin{subnumcases}{}
\label{P1}
P_1'=(m+E-V)P_2,\\
\label{P2}
P_2'=(m-E+V)P_1+\cfrac{2k_d}{r}P_2.
\end{subnumcases}
Clearly, $P_1'\le 0$ which is equivalent to the lemma's third inequality. Near the origin $P_2$ behaves as $c_2r^{\sqrt{k_d^2-f_0^2}+k_d}$, thus $P_2'\ge 0$ near $0$. At infinity (\ref{P2}) becomes $P_2'=(m-E)P_1$, so $P_2'\ge 0$ near infinity. Let us assume that $P_2'<0$ on some $(r_1,\ r_2)$, so $P_2(r_1)=P_2(r_2)=0$. Then by Rolle Theorem, there exists $r_c$ such that $P_2''(r_c)=0$, which corresponds to 
\begin{equation*}
(m-E+V)P_1'+\left(V'+\cfrac{V}{r}\right)P_1+(2k_d-1)\cfrac{P_2'}{r}+(m-E)\cfrac{P_1}{r}=0
\quad\text{at}\quad r=r_c.
\end{equation*}
Since $V=-\cfrac{f}{r}$, the expression $V'+\cfrac{V}{r}=-\cfrac{f'}{r}$ is nonnegative according to $(\rmn{5})$. Therefore in the above expression the first two terms are nonnegative and the last two are strictly positive, which observation reveals a contradiction. Therefore $P_2'\ge 0$ on $r\in[0,\ \infty)$ and this is equivlent to the lemma's last inequality.   

\hfill $\Box$

\medskip

Now we state and prove the refined comparison theorem for a special class of unbounded potentials:

\medskip

\noindent{\bf Theorem 5:} ~~{\it The potential $V=-\cfrac{f}{r}$, where $f$ satisfies $(\rmn{4})$--$(\rmn{6})$, has $\psi_{1i}r^{|k_d|}$ and $\psi_{2i}r^{|k_d|}$--weighted areas, if 
\begin{equation}\label{th7}
\rho_1(r)=\int_0^r (V_b(t)-V_a(t))\psi_{1i}(t)t^{|k_d|}dt\ge 0 \quad \text{and} \quad
\rho_2(r)=\int_0^r (V_b(t)-V_a(t))(-\psi_{2i}(t))t^{|k_d|}dt\ge 0, 
\end{equation}
$r\in [0,\ \infty)$, where $i$ is either $a$ or $b$, then we have $E_a\le E_b$.} 

\medskip

\noindent{\bf Proof:} We prove the theorem for $i=a$; for $i=b$, the proof is the same. As in sec. A we integrate the right side of (\ref{expr5}) by parts to obtain 
\begin{equation*}
\int_0^\infty(V_b-V_a) (\psi_{1a}\psi_{1b} + \psi_{2a}\psi_{2b})dr=
-\int_0^\infty \left[\eta_1\left(\frac{\psi_{1a}}{r^{|k_d|}}\right)' + \eta_2\left(\frac{\psi_{2a}}{r^{|k_d|}}\right)'\right]dr,
\end{equation*}
where $\eta_1(r)$ and $\eta_2(r)$ are defined by (\ref{th7}) for $i=a$.
Then it follows from last two inequalities of the Lemma 2 and (\ref{th7}) that $E_a\le E_b$.

\hfill $\Box$

\medskip

\noindent{\bf Corollary 5:} ~~{\it If the potentials cross over once, say at $r_1$, $V_a\le V_b$ for $r\in [0,\ r_1]$, and
\begin{equation*}\label{concl7}
\rho_1(\infty)=\int_0^\infty (V_b-V_a)\psi_{1i}r^{|k_d|}dr\ge 0 \quad \text{and} \quad
\rho_2(\infty)=\int_0^\infty (V_b-V_a)(-\psi_{2i})r^{|k_d|}dr\ge 0, 
\quad i=a\ or\ b,
\end{equation*}
then $E_a\le E_b$. If the potentials cross over twice, say at $r_1$ and $r_2$, $r_1<r_2$, $V_a\le V_b$ for $r\in [0,\ r_1]$, and
\begin{equation*}\label{concl8}
\rho_1(r_2)=\int_0^{r_2} (V_b-V_a)\psi_{1i}r^{|k_d|}dr\ge 0 \quad \text{and} \quad
\rho_2(r_2)=\int_0^{r_2} (V_b-V_a)(-\psi_{2i})r^{|k_d|}dr\ge 0, 
\quad i=a\ or\ b,  
\end{equation*}
then $E_a\le E_b$.} 

\medskip

The above Corollary can be generalized up to $n$ intersections: if $V_a\le V_b$ on $r\in[0,\ r_1]$ and both sequences of absolute areas $\int_{r_i}^{r_{i+1}}|(V_b-V_a)\psi_{1i}r^{|k_d|}|dr$ and $\int_{r_i}^{r_{i+1}}|(V_b-V_a)\psi_{2i}r^{|k_d|}|dr$, $i=a$ or $b$, are nonincreasing (if $n$ is odd then we assume $\int_{r_{n-1}}^{r_n}|(V_b-V_a)\psi_{1i}r^{|k_d|}|dr\ge\int_{r_n}^\infty|(V_b-V_a)
\psi_{1i}r^{|k_d|}|dr$ and $\int_{r_{n-1}}^{r_n}|(V_b-V_a)\psi_{2i}r^{|k_d|}|dr\ge\int_{r_n}^\infty|(V_b-V_a)
\psi_{2i}r^{|k_d|}|dr$), then $\eta_1(r)\ge 0$ and $\eta_2(r)\ge 0$ for $r\in[0,\ \infty)$, thus $E_a\le E_b$.

%%%%%%%%%%%%%%%%%%%%%%%%%%%%%%%%%%%%%%%%%%%%%%%%%%%%%%%%%%%%%%%%%%%%%%%%%%%%%%%%%%%%%%%%
\subsubsection*{An example}
%%%%%%%%%%%%%%%%%%%%%%%%%%%%%%%%%%%%%%%%%%%%%%%%%%%%%%%%%%%%%%%%%%%%%%%%%%%%%%%%%%%%%%%%
Here we will demonstrate first part of Corollary 5, i. e. the case of one intersection. For the comparison potentials we choose the Hulth\'{e}n potential $V_a$ and the Coulomb potential $V_b$:  
\begin{equation*}
V_a=-\frac{\alpha}{e^{ar}-1} \qquad \text{and} \qquad 
V_b=-\frac{\beta}{r}.
\end{equation*}
The above potentials intersect at exactly one point for $\alpha=1/5$, $a=3/10$, and $\beta=0.508$; Figure 6, left graph. We can see that $V_a\le V_b$ before the intersection point. If $\rho_1(\infty)$ and $\rho_2(\infty)$ are both nonnegative, according to Corollary 5, we have $E_a\le E_b$.

The solutions of the Dirac Coulomb problem are well known. In three dimensional ground state for $j=1/2$, $\tau=-1$, and $m=2$ the eigenvalue is \cite{Gr}
\begin{equation*}
E_b=2\gamma
\end{equation*}
and the wave functions are
\begin{equation*}
\left.
\begin{array}{ll}
\psi_{1b} \\
\psi_{2b} 
\end{array} \right\}=\pm
\sqrt{\cfrac{2\beta(1\pm\gamma)}{\Gamma(2\gamma+1)}}\cfrac{(4\beta r)^\gamma}{e^{2\beta r}},
\end{equation*}
where $\gamma=\sqrt{1-\beta^2}$, and $\Gamma$ is the gamma function (the wave functions are plotted on Figure 6, right graph). Then $\rho_1(\infty)=0.00113$ and $\rho_2(\infty)=0.00031$ and, according to accurate numerical calculation, $E_a=1.58604\le E_b=1.72271$.
\begin{figure}[ht]
\begin{center}$
\begin{array}{cc}
\includegraphics[width=2.5in, angle=-90]{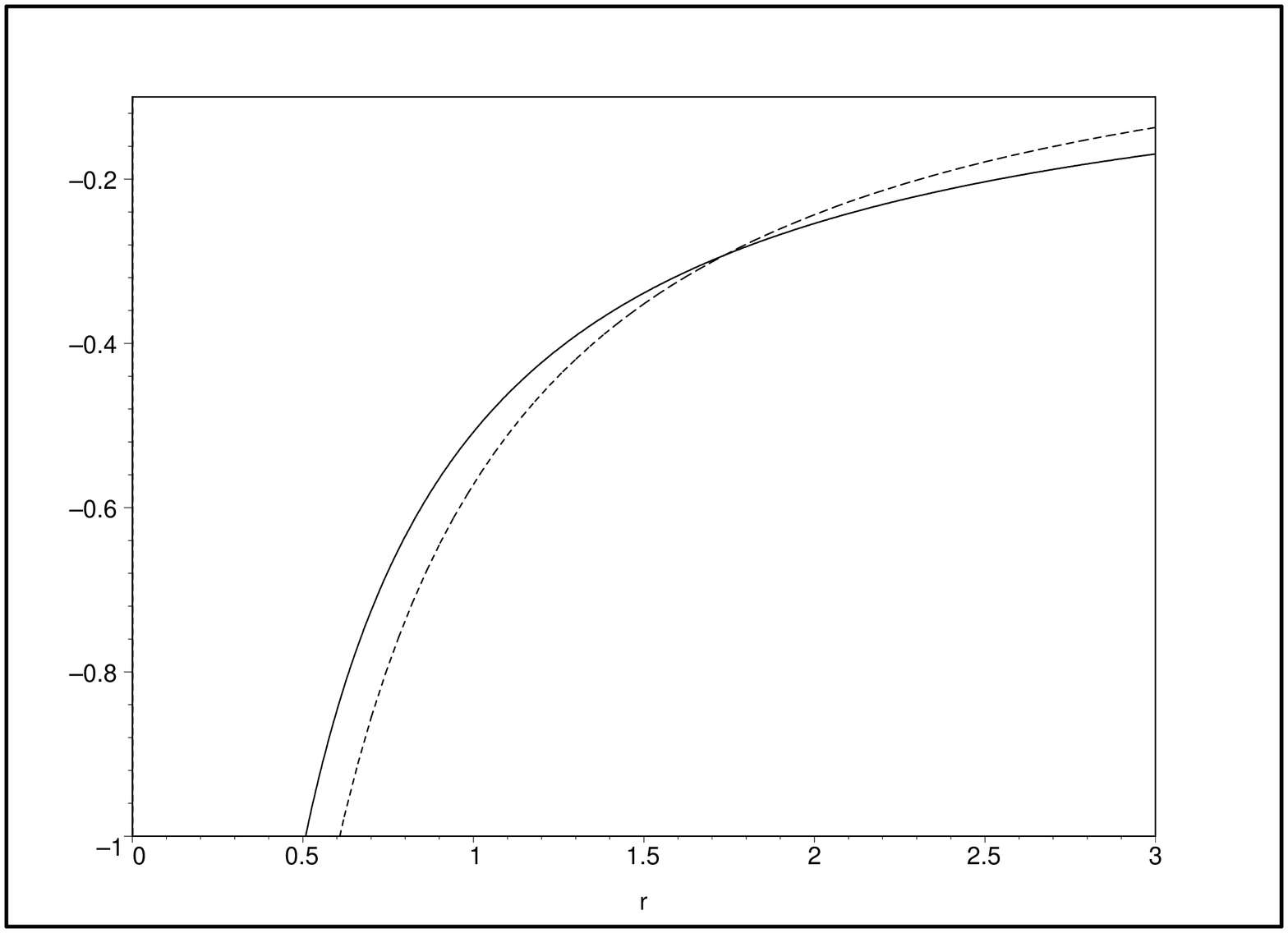} & 
\includegraphics[width=2.5in, angle=-90]{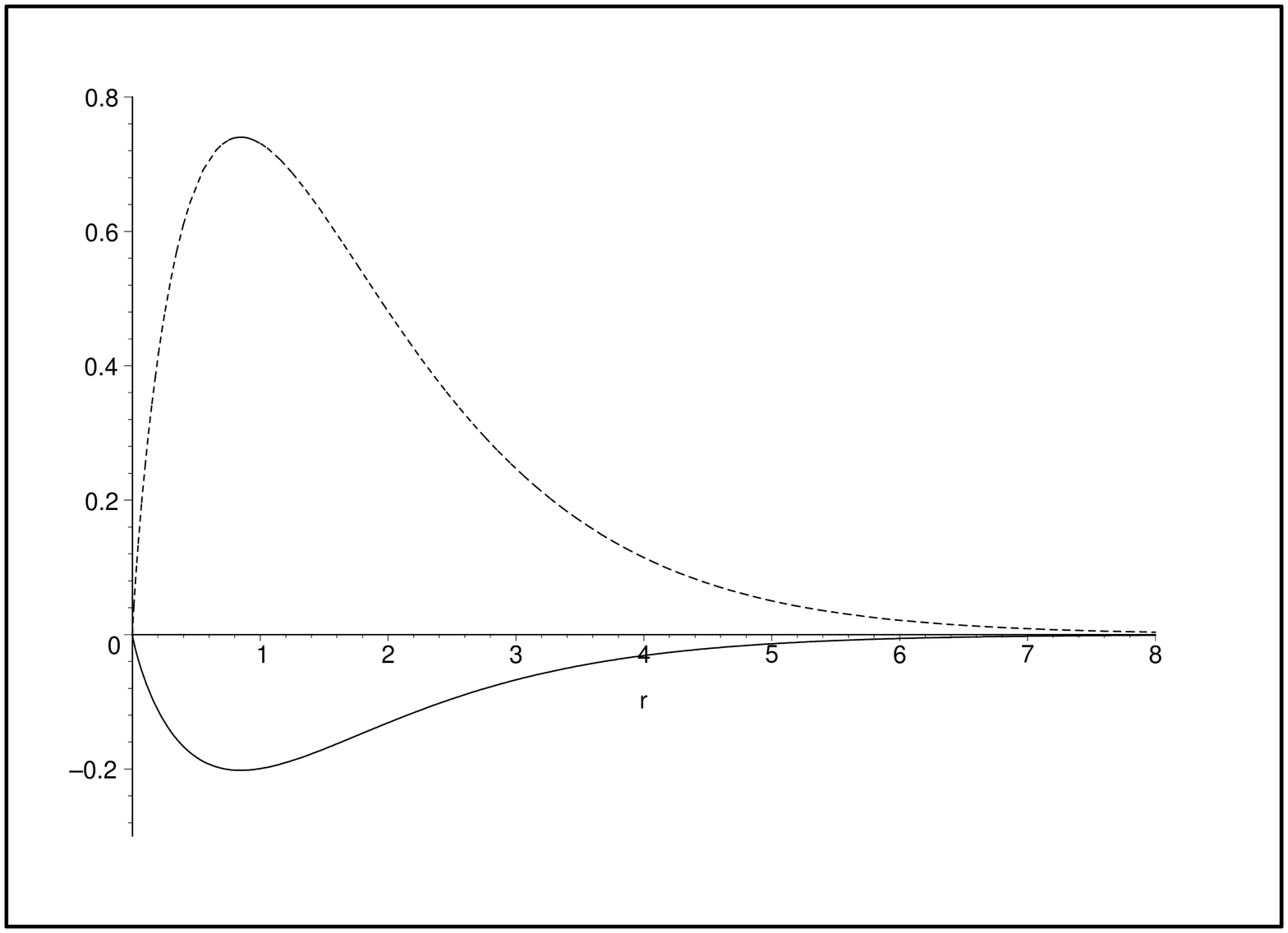}
\end{array}$
\end{center}
\caption{Left graph: The Hulth\'en potential $V_a$ dotted line and the Coulomb potential $V_b$ full line. Right graph: Upper component of the Dirac spinor $\psi_{1b}$ dotted line and lower component $\psi_{2b}$ full line.}
\end{figure}
\begin{figure}[ht] % F7
\begin{center}$
\begin{array}{cc}
\includegraphics[width=2.5in, angle=-90]{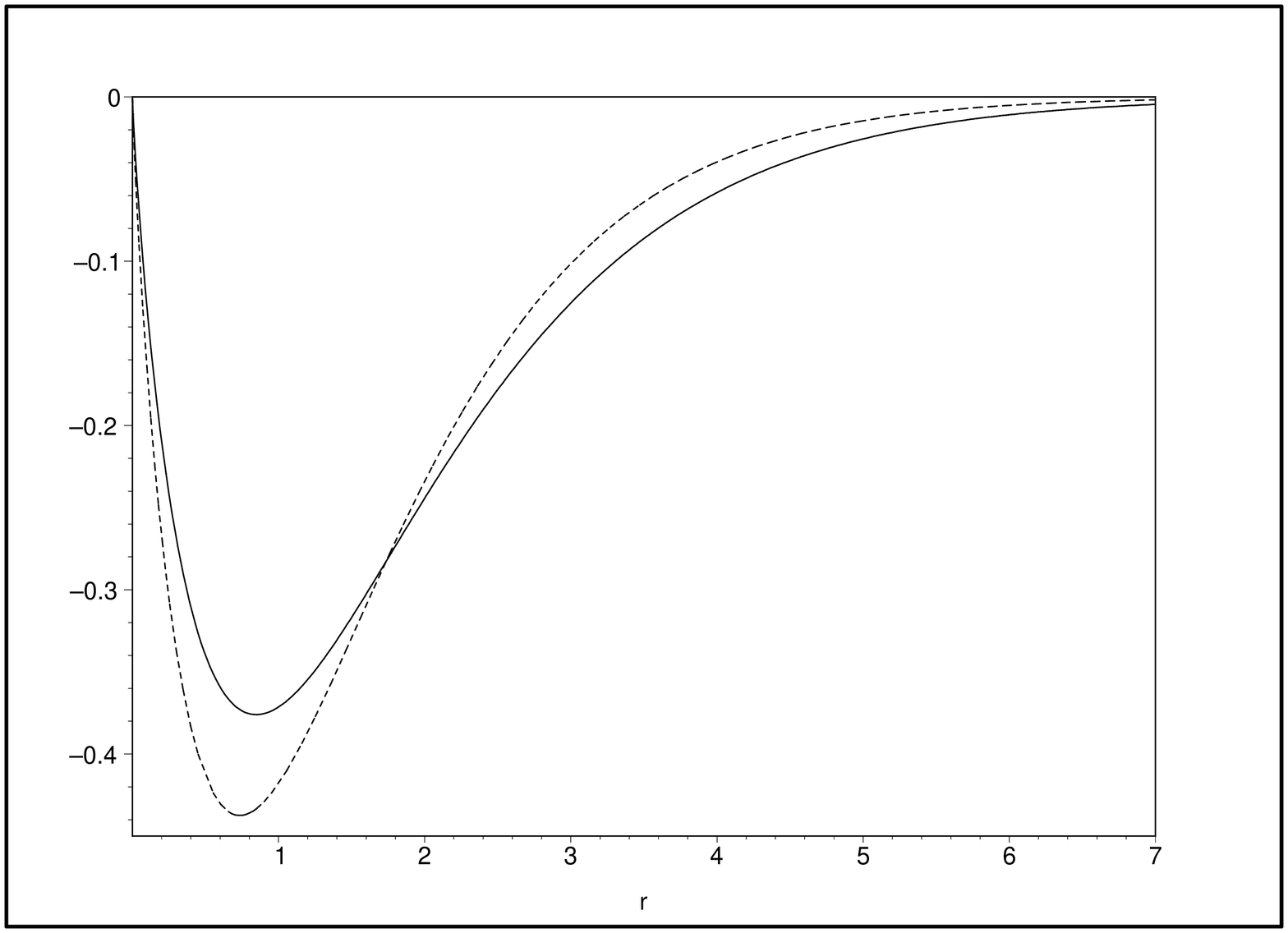} & 
\includegraphics[width=2.5in, angle=-90]{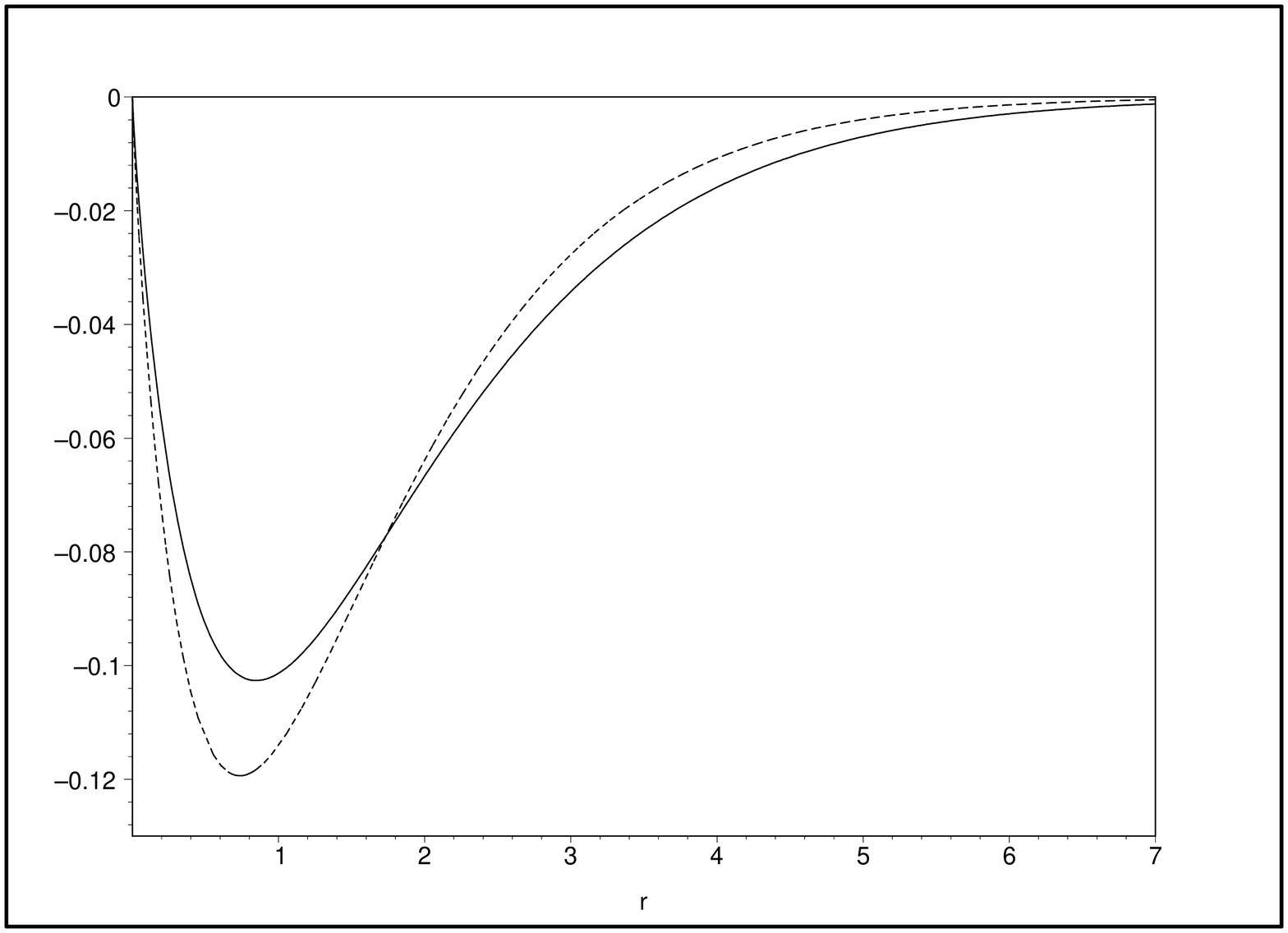}
\end{array}$
\end{center}
\caption{Left graph: Function $V_a\psi_{1b}r$ dotted line and $V_b\psi_{1b}r$ full line. Right graph: Function $V_a(-\psi_{2b})r$ dotted line and $V_b(-\psi_{2b})r$ full line.}
\end{figure}
\begin{figure}[ht] %F8
\begin{center}$
\begin{array}{cc}
\includegraphics[width=2.5in, angle=-90]{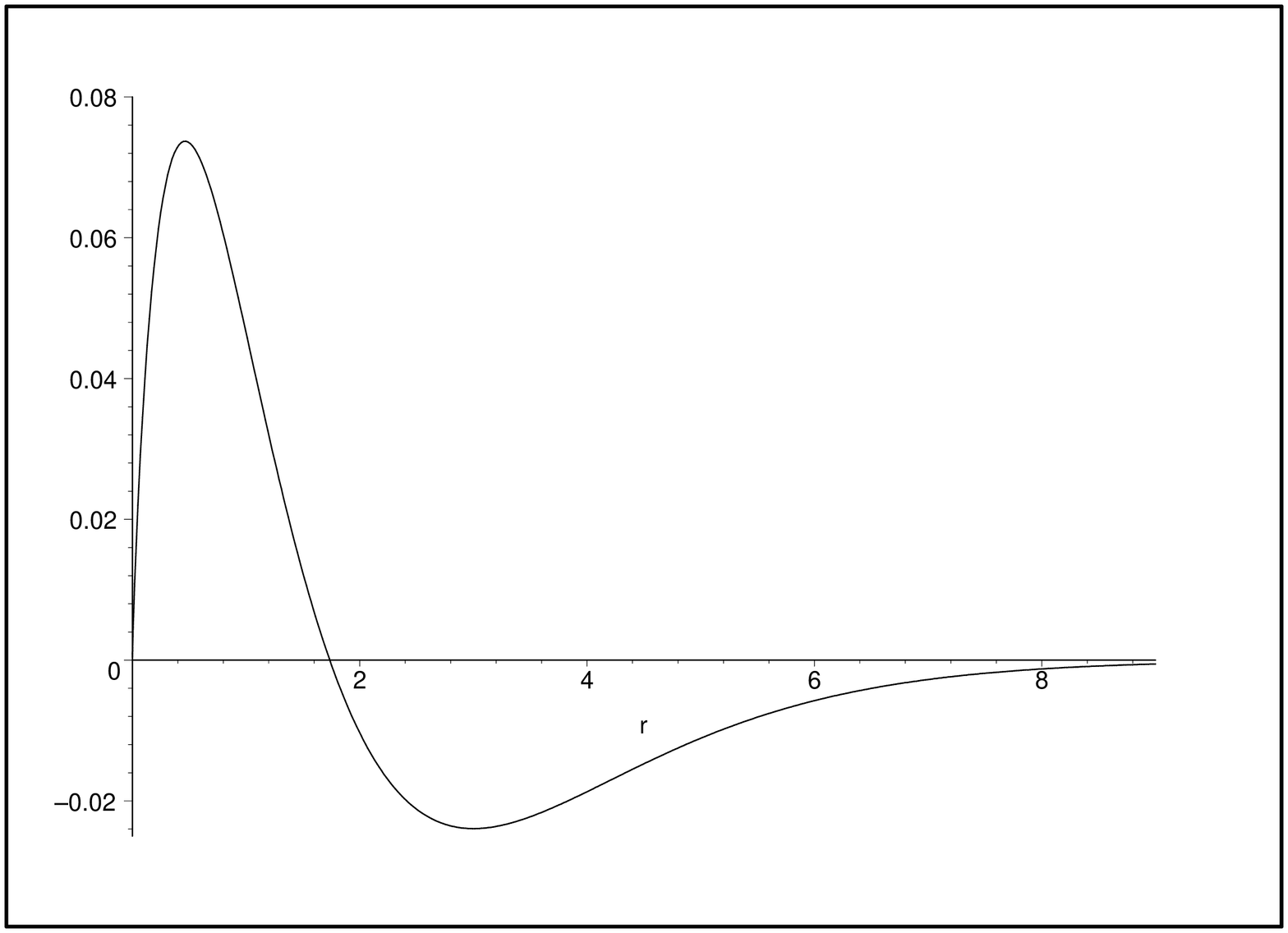} & 
\includegraphics[width=2.5in, angle=-90]{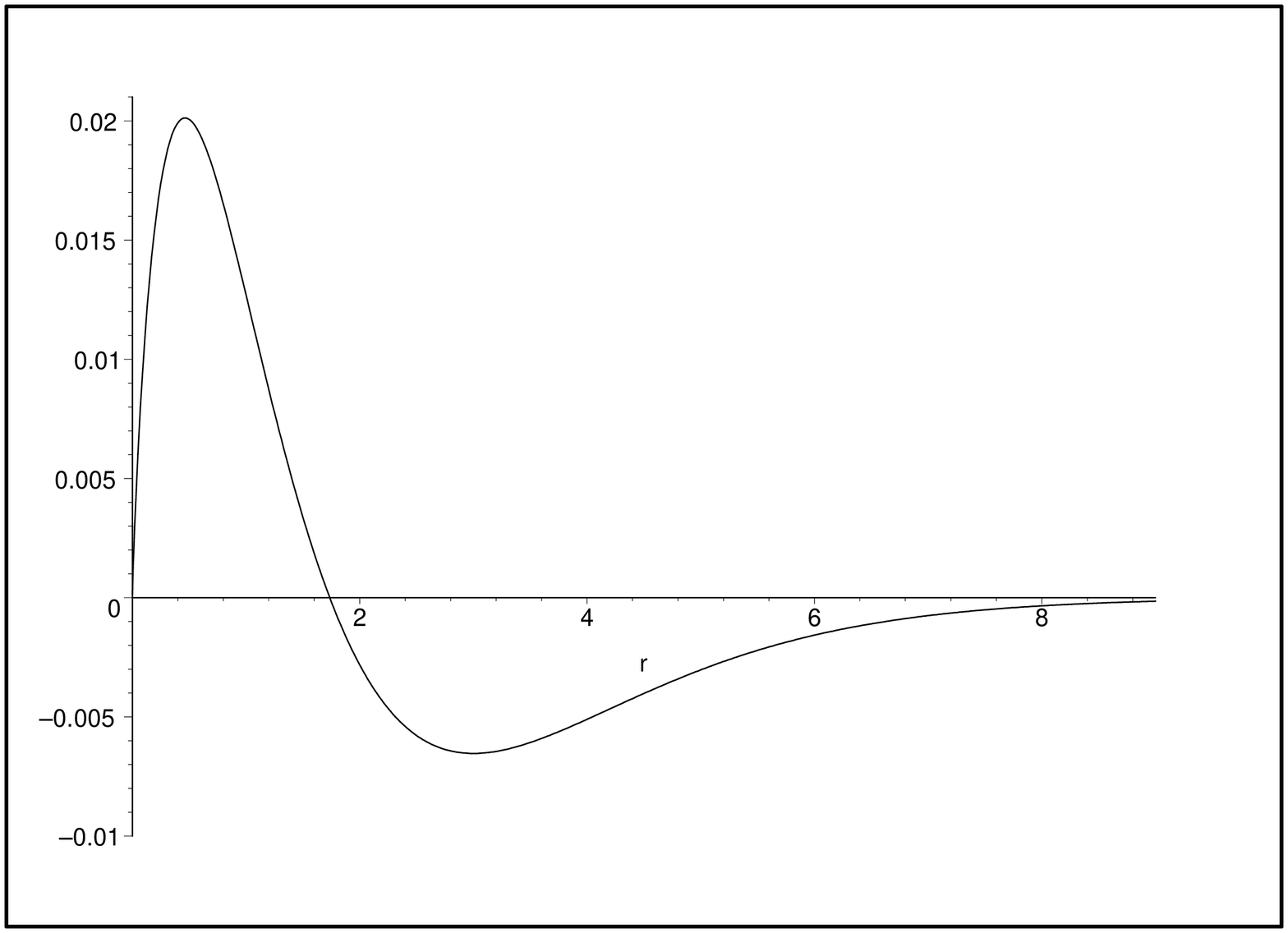}
\end{array}$
\end{center}
\caption{Left graph: The graph of the integrand $(V_b-V_a)\psi_{1b}r$. Right graph: The graph of the integrand $(V_b-V_a)(-\psi_{2b})r$.}
\end{figure}

%%%%%%%%%%%%%%%%%%%%%%%%%%%%%%%%%%%%%%%%%%%%%%%%%%%%%%%%%%%%%%%%%%%%%%%%%%%%%%%%%%%%%%%%
\subsection{Potentials less singular than Coulomb.}
%%%%%%%%%%%%%%%%%%%%%%%%%%%%%%%%%%%%%%%%%%%%%%%%%%%%%%%%%%%%%%%%%%%%%%%%%%%%%%%%%%%%%%%%
We characterize this class of potentials in the following way:
\begin{eqnarray*}
&(\rmn{7})& V\ \text{is nonpositive and unbounded, i.e.}\ V\le 0\ \text{and}\ \lim\limits_{r\to 0^+}V=-\infty;\\
\vspace{9mm}
&(\rmn{8})& V\ \text{is attractive, that is monotone nondecreasing on} \  [0, \infty)\ \text{so}\  V'\ge 0;\\
\vspace{9mm}
&(i\times)& V\ \text{vanishes at infinity, thus}\ \lim_{r\to\infty}V=0;\\
\vspace{9mm}
&(\times)& V\ \text{is less singular than Coulomb potential, i.e.}\ \lim\limits_{r\to 0^+}rV=0.
\end{eqnarray*}
Examples of such potentials are: $V=-\cfrac{v}{r^q}$, $V=-\cfrac{v}{r^q+r^b}$, and $V=-\cfrac{v}{r^qe^{br}}$, where $q\in(0,\ 1)$, $v$ and $b$ are positive constants. Then, as before, we first prove a lemma, namely

\medskip

\noindent{\bf Lemma 3:} ~~{\it the Dirac radial spinor components $\psi_1$ and $\psi_2$ at the bottom of an angular--momentum subspace labelled by $j$, which satisfy (\ref{dcel1})--(\ref{dcer1}), for the potential $V$, which satisfies $(i)$--$(iv)$, are such that
\begin{equation*}
\left(\frac{\psi_1}{r^{|k_d|}}\right)'\le 0 \quad \text{and} \quad
\left(\frac{\psi_2}{r^{|k_d|+1}}\right)'\ge 0, \quad r\in[0, \ \infty).
\end{equation*}} 

\medskip

\noindent{\bf Proof:} Near the origin the above class of the potentials can be approximated by $V=-\cfrac{v}{r^q}$, then the system (\ref{dcel1})--(\ref{dcer1}) after some rearrangements becomes: 
\begin{subnumcases}{}
\psi_1''=\psi_1\left(\frac{k_d(k_d+1-q)}{r^2}-\frac{v^2}{r^{2q}}\right)-
\psi_1'\frac{q}{r},\\
\psi_2''=\psi_2\left(\frac{k_d(k_d-1+q)}{r^2}-\frac{v^2}{r^{2q}}\right)-
\psi_2'\frac{q}{r}.
\end{subnumcases}
Solutions of these equations are given in terms of Bessel functions. Therefore for small $r$ we can approximate them by simple powers
\begin{subnumcases}{}
\label{powl'}
\psi_1=c_1r^{q_1},\\
\label{powr'}
\psi_2=c_2r^{q_2},
\end{subnumcases}
where $c_1$ and $c_2$ are constants of integration and parameters $q_1$ and $q_2$ are positive since both wave functions must vanish at the origin. Then, following the proof of Lemma 1, we find $q_1=-k_d$, $q_2=-k_d-q+1$, and $\cfrac{c_1}{c_2}=-\cfrac{1-q-2k_d}{v}<0$. Thus, near the origin radial wave functions behave as
\begin{subnumcases}{}
\nonumber
\psi_1=c_1r^{-k_d},\\
\nonumber
\psi_2=c_2r^{-k_d-q+1}.
\end{subnumcases}

Now let us make the following substitution $\psi_1=r^{-k_d}R_1$ and $\psi_2=r^{-k_d+1}R_2$, then the system of equations (\ref{dcel1})--(\ref{dcer1}) becomes
\begin{subnumcases}{}
\label{dcel2'}
R_1'=(m+E-V)rR_2,\\
\label{dcer2'}
R_2'=(m-E+V)\frac{R_1}{r}+\frac{2k_d-1}{r}R_2.
\end{subnumcases}
According to $(\rmn{7})$ and (\ref{dcel2'}), $R_1'\le 0$ which is equivalent to the lemma's first inequality. Near $0$ the function $R_2$ behaves as $c_2r^{-q}$, thus $R_2'\ge 0$ near $0$. Equations (\ref{dcel2'})--(\ref{dcer2'}) are exactly the same as (\ref{dcel2})--(\ref{dcer2}), thus it can be proved by contradiction that $R_2'\ge 0$ on $r\in[0, \ \infty)$ which is the same as the lemma's second inequality.   

\hfill $\Box$

\medskip

Since Lemma 1 and Lemma 3 have the same conclusions, the following theorem along with the corollary can be stated and proved as in the bounded case:

\medskip

\noindent{\bf Theorem 6:} ~~{\it The potential $V$ satisfies $(\rmn{7})$--$(\rmn{10})$, has $\psi_{1i}r^{|k_d|}$ and $\psi_{2i}r^{|k_d|+1}$--weighted areas, if 
\begin{equation}\label{th6'}
\mu_1(r)=\int_0^r (V_b(t)-V_a(t))\psi_{1i}(t)t^{|k_d|}dt\ge 0 \quad \text{and} \quad
\mu_2(r)=\int_0^r (V_b(t)-V_a(t))(-\psi_{2i}(t))t^{|k_d|+1}dt\ge 0, 
\end{equation}
$r\in [0,\ \infty)$, where $i$ is either $a$ or $b$, then we have $E_a\le E_b$.} 

\medskip

\noindent{\bf Corollary 6:} ~~{\it If the potentials cross over once, say at $r_1$, $V_a\le V_b$ for $r\in [0,\ r_1]$, and
\begin{equation*}
\mu_1(\infty)=\int_0^\infty (V_b-V_a)\psi_{1i}r^{|k_d|}dr\ge 0 \quad \text{and} \quad
\mu_2(\infty)=\int_0^\infty (V_b-V_a)(-\psi_{2i})r^{|k_d|+1}dr\ge 0, 
\quad i=a\ or\ b,
\end{equation*}
then $E_a\le E_b$. If the potentials cross over twice, say at $r_1$ and $r_2$, $r_1<r_2$, $V_a\le V_b$ for $r\in [0,\ r_1]$, and
\begin{equation*}
\mu_1(r_2)=\int_0^{r_2} (V_b-V_a)\psi_{1i}r^{|k_d|}dr\ge 0 \quad \text{and} \quad
\mu_2(r_2)=\int_0^{r_2} (V_b-V_a)(-\psi_{2i})r^{|k_d|+1}dr\ge 0, 
\quad i=a\ or\ b,  
\end{equation*}
then $E_a\le E_b$.} 

\medskip

As before Corollary 6 can be generalized to the case of $n$ intersections.

%%%%%%%%%%%%%%%%%%%%%%%%%%%%%%%%%%%%%%%%%%%%%%%%%%%%%%%%%%%%%%%%%%%%%%%%%%%%%%%%%%%%%%%%
\subsection{General refined comparison theorem}
%%%%%%%%%%%%%%%%%%%%%%%%%%%%%%%%%%%%%%%%%%%%%%%%%%%%%%%%%%%%%%%%%%%%%%%%%%%%%%%%%%%%%%%%
In this section we shall construct the refined comparison theorem which combines all the above cases, i.e. bounded, unbounded, and less singular than Coulomb potentials. First, we state the following lemma:

\medskip

\noindent{\bf Lemma 4:} ~~{\it the Dirac radial spinor components $\psi_1$ and $\psi_2$ at the bottom of an angular--momentum subspace labelled by $j$, which satisfy (\ref{dcel1})--(\ref{dcer1}), for the bounded potential (satisfies (\rmn{1})--(\rmn{3})), or for the unbounded potential (satisfies (\rmn{4})--(\rmn{6})), or for the potential less singular than Coulomb (satisfies (\rmn{7})--($\times$)) are such that
\begin{equation*}
\left(\frac{\psi_1}{r^{|k_d|}}\right)'\le 0 \quad \text{and} \quad
\left(\frac{\psi_2}{r^{|k_d|+1}}\right)'\ge 0, \quad r\in[0, \ \infty).
\end{equation*}} 

\medskip

\noindent{\bf Proof:} The proof for the bounded and less singular than Coulomb potentials follows from Lemma 1 and 3 respectively, so we have to prove the Lemma for the unbounded case only. 

Let us consider equations (\ref{dcel1})--(\ref{dcer1}) with substitution $\psi_1=r^{-k_d}G_1$ and $\psi_2=r^{-k_d+1}G_2$:
\begin{subnumcases}{}
\label{dcel21}
G_1'=(m+E-V)rG_2,\\
\label{dcer21}
G_2'=(m-E+V)\frac{G_1}{r}+\frac{2k_d-1}{r}G_2.
\end{subnumcases}
Then the lemma's first inequality follows from (\ref{dcel21}). We know from the proof of Lemma 2 that $G_2$ behaves near the origin as $c_2r^{\sqrt{k_d^2-f_0^2}+k_d-1}$, where the constant $c_2<0$, therefore $G_2'\ge 0$ near $0$. Since $(m-E+V)\ge 0$ near infinity, then according to (\ref{dcer21}), function $G_2'\ge 0$ near infinity. Finally, it can be shown by contradiction that $G_2'\ge 0$ on $r\in[0,\ \infty)$, which ends the proof.

\hfill $\Box$

Using the above Lemma, the general refined comparison theorem can be proved:

\medskip

\noindent{\bf Theorem 7:} ~~{\it The potential $V$ satisfies either (\rmn{1})--(\rmn{3}), or (\rmn{4})--(\rmn{6}), or (\rmn{7})--($\times$), has $\psi_{1i}r^{|k_d|}$ and $\psi_{2i}r^{|k_d|+1}$--weighted areas, if 
\begin{equation}\label{th8}
\zeta_1(r)=\int_0^r (V_b(t)-V_a(t))\psi_{1i}(t)t^{|k_d|}dt\ge 0 \quad \text{and} \quad
\zeta_2(r)=\int_0^r (V_b(t)-V_a(t))(-\psi_{2i}(t))t^{|k_d|+1}dt\ge 0, 
\end{equation}
$r\in [0,\ \infty)$, where $i$ is either $a$ or $b$, then we have $E_a\le E_b$.} 

\medskip

\noindent{\bf Corollary 7:} ~~{\it If the potentials cross over once, say at $r_1$, $V_a\le V_b$ for $r\in [0,\ r_1]$, and
\begin{equation*}
\zeta_1(\infty)=\int_0^\infty (V_b-V_a)\psi_{1i}r^{|k_d|}dr\ge 0 \quad \text{and} \quad
\zeta_2(\infty)=\int_0^\infty (V_b-V_a)(-\psi_{2i})r^{|k_d|+1}dr\ge 0, 
\quad i=a\ or\ b,
\end{equation*}
then $E_a\le E_b$. If the potentials cross over twice, say at $r_1$ and $r_2$, $r_1<r_2$, $V_a\le V_b$ for $r\in [0,\ r_1]$, and
\begin{equation*}
\zeta_1(r_2)=\int_0^{r_2} (V_b-V_a)\psi_{1i}r^{|k_d|}dr\ge 0 \quad \text{and} \quad
\zeta_2(r_2)=\int_0^{r_2} (V_b-V_a)(-\psi_{2i})r^{|k_d|+1}dr\ge 0, 
\quad i=a\ or\ b,  
\end{equation*}
then $E_a\le E_b$.} 

\medskip

As before Corollary 7 can be generalized to the case of $n$ intersections.

%%%%%%%%%%%%%%%%%%%%%%%%%%%%%%%%%%%%%%%%%%%%%%%%%%%%%%%%%%%%%%%%%%%%%%%%%%%%%%%%%%%%%%%%
\subsubsection*{An example}
%%%%%%%%%%%%%%%%%%%%%%%%%%%%%%%%%%%%%%%%%%%%%%%%%%%%%%%%%%%%%%%%%%%%%%%%%%%%%%%%%%%%%%%%
In that section we will demonstrate the second part of Corollary 7, using unbounded Coulomb potential $V_a$ and bounded Sech--squared potential $V_b$, which has been known under several different names since early days of quantum mechanics such as the P\"{o}schl–-Teller potential \cite{PT} or the Eckart potential \cite{Eckart}:
\begin{equation*}
V_a=-\frac{\alpha}{r} \qquad \text{and} \qquad 
V_b=-\beta \sech^2br.
\end{equation*}
If $\alpha=0.579$, $\beta=0.3$, and $b=0.2$ then $V_a$ and $V_a$ intersect at exactly two points $r_1=2.41742$ and $r_2=5.66301$; see Figure 9. According to \cite{Gr}, in $d=3$ dimensions for $j=1/2$, $\tau=-1$, and $m=1$ the wave functions and the eigenvalue are:
\begin{equation*}
\left.
\begin{array}{ll}
\psi_{1a} \\
\psi_{2a} 
\end{array} \right\}=\pm
\sqrt{\cfrac{\alpha(1\pm\gamma)}{\Gamma(2\gamma+1)}}\cfrac{(2\alpha r)^\gamma}{e^{\alpha r}},
 \qquad \text{and} \qquad 
E_a=\gamma,
\end{equation*}
where $\gamma=\sqrt{1-\alpha^2}$, and $\Gamma$ is the gamma function. Then direct calculation shows that $\zeta_1(r_2)=0.18778$ and $\zeta_2(r_2)=0.00084$. Thus Corollary 7 implies that $E_a\le E_b$, which agrees with the accurate numerical energy eigenvalies: $E_a=0.81533\le E_b=0.88318$.
\begin{figure} %F9
\centering{\includegraphics[height=13cm,width=8cm,angle=-90]{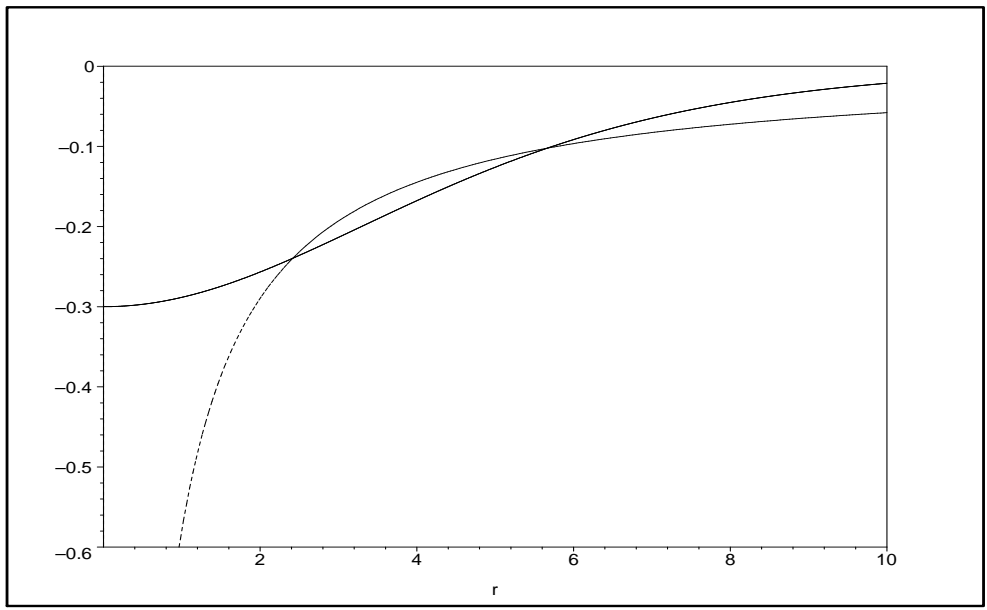}}
\caption{The Coulomb potential $V_a$ dotted line and the Sech--squared potential $V_b$ full line.}
\end{figure}
\begin{figure}[ht] % F10
\begin{center}$
\begin{array}{cc}
\includegraphics[width=2.5in, angle=-90]{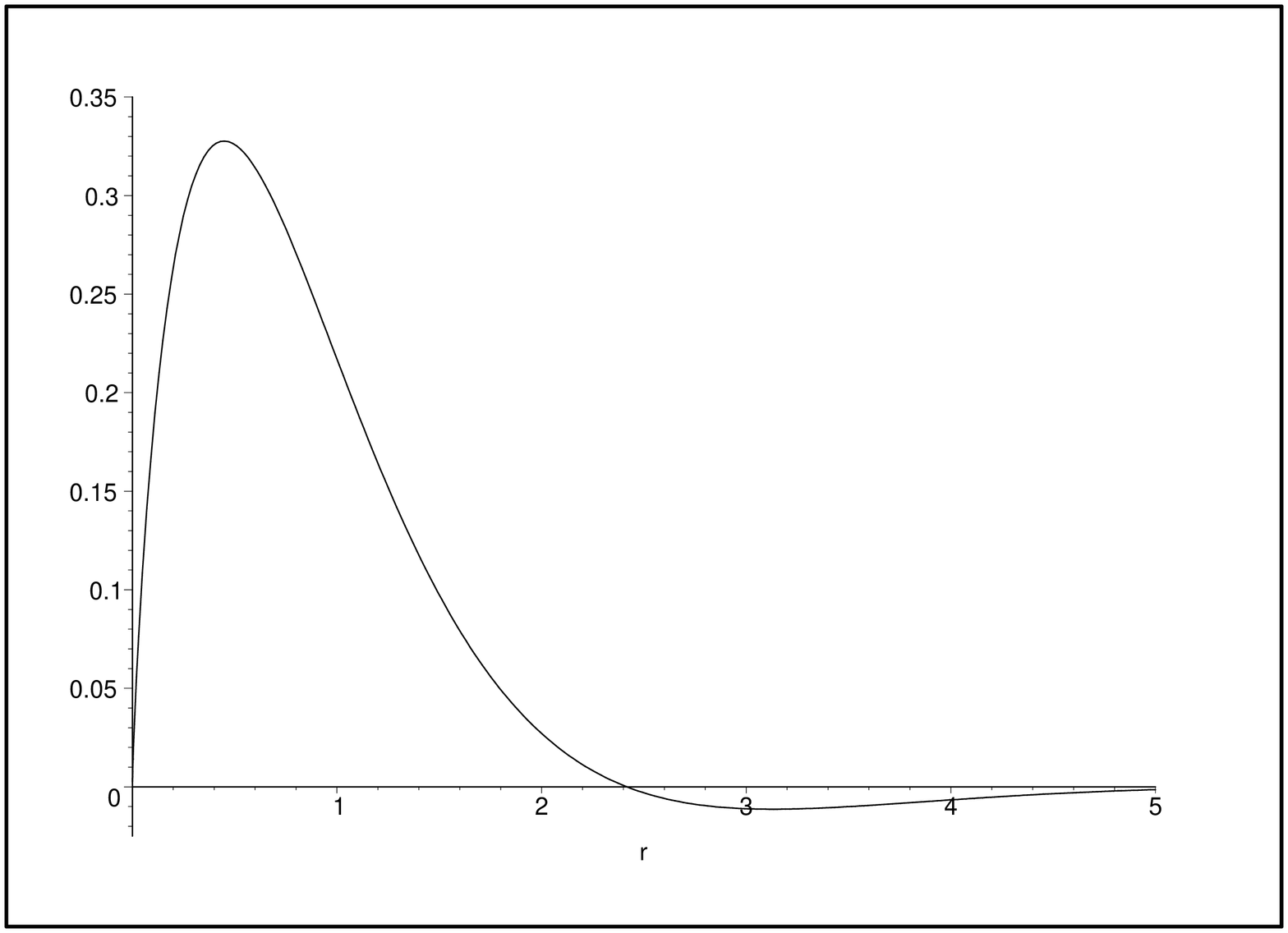} & 
\includegraphics[width=2.5in, angle=-90]{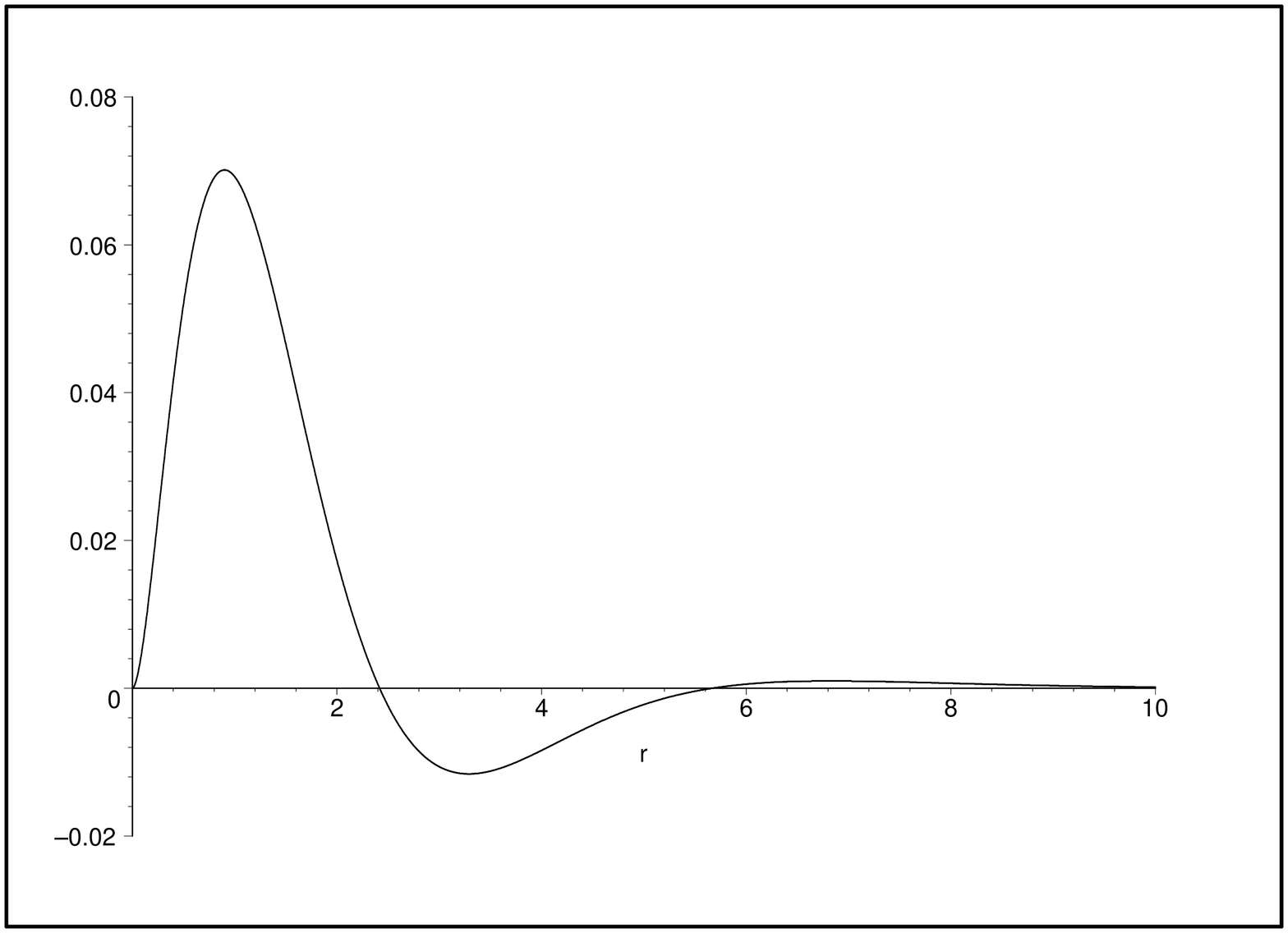}
\end{array}$
\end{center}
\caption{Left graph: The graph of the integrand $(V_b-V_a)\psi_{1b}r$. Right graph: The graph of the integrand $(V_b-V_a)(-\psi_{2b})r^2$.}
\end{figure}

%%%%%%%%%%%%%%%%%%%%%%%%%%%%%%%%%%%%%%%%%%%%%%%%%%%%%%%%%%%%%%%%%%%%%%%%%%%%%%%%%%%%%%%%
%%%%%%%%%%%%%%%%%%%%%%%%%%%%%%%%%%%%%%%%%%%%%%%%%%%%%%%%%%%%%%%%%%%%%%%%%%%%%%%%%%%%%%%%
\section{Conclusion}
%%%%%%%%%%%%%%%%%%%%%%%%%%%%%%%%%%%%%%%%%%%%%%%%%%%%%%%%%%%%%%%%%%%%%%%%%%%%%%%%%%%%%%%%
%%%%%%%%%%%%%%%%%%%%%%%%%%%%%%%%%%%%%%%%%%%%%%%%%%%%%%%%%%%%%%%%%%%%%%%%%%%%%%%%%%%%%%%%
In this paper we have rederived the earlier comparison theorems in $d$ dimensions and then refined these theorems by allowing the comparison potentials to crossover in a controlled manner. Because of the different form of the Dirac coupled equations in $d=1$ and $d>1$ dimensions we have studied these cases separately. We have shown that in one dimension the condition $V_a\le V_b$, which leads to the spectral ordering $E_a\le E_b$, can be replaced by weaker condition $U_a\le U_b$, where $U_i(x)=\int_0^x V_i(t)dt$ $i=a$ or $b$. Since the potential cross--over conditions depend on the detailed behaviour of the radial wave--functions components, we have found it best in $d>1$ dimensions to establish first a separate refined comparison theorem for each of a number of interesting classes of potential: Theorems 3 and 4 for bounded potentials, Theorem 5 for unbounded potentials, and Theorem 6 for the class of potentials less singular than Coulomb. Finally, we have summarized the refined comparison results in Theorem 7, which combines all of the above types of potentials and states that if $U_{1a}\le U_{1b}$ and $U_{2a}\le U_{2b}$ then $E_a\le E_b$, where $U_{1i}(r)=\int_0^r V_i(t)\psi_{1j}(t)t^{|k_d|} dt$ and $U_{2i}(r)=\int_0^r V_i(t)(-\psi_{2j}(t))t^{|k_d|+1} dt$, where $i,\ j=a$ or $b$.

For practical reasons we have also established weaker sufficient conditions, as corollaries, which guarantee in simple ways that the comparison potentials crossover so as to imply definite spectral ordering. These results are illustrated by some specific examples.

%%%%%%%%%%%%%%%%%%%%%%%%%%%%%%%%%%%%%%%%%%%%%%%%%%%%%%%%%%%%%%%%%%%%%%%%%%%%%%%%%%%%%%%%
%%%%%%%%%%%%%%%%%%%%%%%%%%%%%%%%%%%%%%%%%%%%%%%%%%%%%%%%%%%%%%%%%%%%%%%%%%%%%%%%%%%%%%%%
\section{Acknowledgments}
%%%%%%%%%%%%%%%%%%%%%%%%%%%%%%%%%%%%%%%%%%%%%%%%%%%%%%%%%%%%%%%%%%%%%%%%%%%%%%%%%%%%%%%%
%%%%%%%%%%%%%%%%%%%%%%%%%%%%%%%%%%%%%%%%%%%%%%%%%%%%%%%%%%%%%%%%%%%%%%%%%%%%%%%%%%%%%%%%
One of us (RLH) gratefully acknowledges partial financial support
of this research under Grant No.\ GP3438 from the Natural Sciences
and Engineering Research Council of Canada.\medskip

%%%%%%%%%%%%%%%%%%%%%%%%%%%%%%%%%%%%%%%%%%%%%%%%%%%%%%%%%%%%%%%%%%%%%%%%%%%%%%%%%%%%%%%%
%%%%%%%%%%%%%%%%%%%%%%%%%%%%%%%%%%%%%%%%%%%%%%%%%%%%%%%%%%%%%%%%%%%%%%%%%%%%%%%%%%%%%%%%
%\section*{References}
%%%%%%%%%%%%%%%%%%%%%%%%%%%%%%%%%%%%%%%%%%%%%%%%%%%%%%%%%%%%%%%%%%%%%%%%%%%%%%%%%%%%%%%%
%%%%%%%%%%%%%%%%%%%%%%%%%%%%%%%%%%%%%%%%%%%%%%%%%%%%%%%%%%%%%%%%%%%%%%%%%%%%%%%%%%%%%%%%


\begin{thebibliography}{99}
\bibitem{p75} R. L. Hall, Phys. Rev. Lett. {\bf 83}, 468 (1999).
\bibitem{Reed} M. Reed and B. Simon, {\it Methods of Modern Mathematical Physics IV: Analysis of Operators}, (Academic, New York, 1978). 
\bibitem{Thirring} W. Thirring, {\it A Course in Mathematical Physics 3: Quantum Mechanics of Atoms and Molecules}, (Springer, New York, 1981). 
\bibitem{Fr} J. Franklin and L. Intemann, Phys. Rev. Lett. {\bf 54}, 2068 (1985).
\bibitem{Gold} S. P. Goldman, Phys. Rev. A {\bf 31}, 3541 (1985).
\bibitem{Gr} I. P. Grant and H. M. Quiney, Phys. Rev. A {\bf 62}, 022508 (2000).
\bibitem{chen1} G. Chen, Phys. Rev. A {\bf 71}, 024102 (2005).
\bibitem{chen2} G. Chen, Phys. Rev. A {\bf 72}, 044102 (2005).
\bibitem{monoton2} R. L. Hall and M. D. Aliyu, Phys. Rev. A {\bf 78}, 052115 (2008).
\bibitem{HY} R. L. Hall and \"{O}. Ye\c{s}ilta\c{s}, J. Phys. A: Math. Theor. {\bf 43}, 195303 (2010).
\bibitem{p127} R. L. Hall, Phys. Rev. Lett. {\bf 101}, 090401 (2008).
\bibitem{p134} R. L. Hall, Phys. Rev. A {\bf 81}, 052101 (2010).
\bibitem{Semay} C. Semay, Phys. Rev. A {\bf 83}, 024101 (2011).
\bibitem{HF} H. Hellmann, Acta Physicochim. URSS {\bf 1}, 913 (1935); R. P. Feynmann, Phys. Rev. {\bf 56}, 340 (1939).
\bibitem{Hall7}R. L. Hall, J. Phys. A: Math. Gen. {\bf 25}, 4459 (1992). 
\bibitem{ddimSch} R. L. Hall and Q. D. Katatbeh, J. Phys. A {\bf 35}, 8727 (2002).
\bibitem{Hall44} R. L. Hall and N. Saad, Phys. Lett. A {\bf 237}, 107 (1998).
\bibitem{rose}M.~E.~Rose and R.~R.~Newton, Phys. Rev. {\bf 82}, 470 (1951).
\bibitem{rose_book} M. E. Rose, {\it Relativistic Electron Theory}, (USA, Wiley, 1961)
Nodes of the Radial Functions are discussed on page 166.
\bibitem{nod} R. L. Hall and P. Zorin, Ann. Phys. (Berlin) {\bf 526}, 79 (2014); arXiv:1309.1749v1, (2013).
\bibitem{calog}A. Calogeracos, N. Dombey, and K. Imagawa, Phys. Atom. Nucl. {\bf 59}, 1275 (1996); Yad. Fiz. {\bf 59}, 1331 (1996).
\bibitem{Spectrumd11} E. C. Titchmarsh, Quart. J. Math. Oxford (2) {\bf 13}, 255 (1962).
\bibitem{Spectrumd12} S. Golenia and T. Haugomat, arXiv:1207.3516v1, (2012).
\bibitem{Dombey}N. Dombey, P. Kennedy, and A. Calogeracos, Phys. Rev. Lett. {\bf 85}, 1787 (2000). 
\bibitem{Qiong} Qiong--gui Lin, Eur. Phys. J. D {\bf 7}, 515 (1999).
\bibitem{Flugge}S. Fl\"{u}gge, {\it Practical Quantum Mechanics}, (Springer--Verlag, Berlin, 1999). 
\bibitem{laser1} D. Singh and Y. P. Varshni, Phys. Rev. A {\bf 32}, 619 (1985).
\bibitem{laser2} R. Dutt, U. Mukherji, and Y. P. Varshni, J. Phys. B: At. Mol. Phys. {\bf 18}, 3311 (1985).
\bibitem{Hall3}J. P. Duarte and R. L. Hall, J. Phys. B: At. Mol. Opt. Phys. {\bf 27}, 1021 (1994).
\bibitem{Bjorken} J. D. Bjorken and S. D. Drell, {\it Relativistic Quantum Mechanics}, (McGraw-Hill Book Co., New York, 1964).
\bibitem{Gu} X. Y. Gu, Z. Q. Ma, and S. H. Dong, Int. J. Mod. Phys. E {\bf 11}, 335 (2002).
\bibitem{Dong} S. H. Dong, J. Phys. A: Math. Theor. {\bf 36}, 4977 (2003).  
\bibitem{jiang}Y. Jiang, J. Phys. A {\bf 38}, 1157 (2005).
\bibitem{salazar}M. Salazar-Ramirez, D. Martinez, R. D. Mota, and V. D. Granados, EPL {\bf 95}, 60002 (2011). 
\bibitem{yasuk}F. Yasuk and M. K. Bahar, Phys. Scr. {\bf 85}, 045004 (2012).
\bibitem{messiah}A. Messiah, {\it Quantum Mechanics}, (North Holland, Amsterdam, 1962). The Dirac equation for central fields is discussed on page 928.
\bibitem{greiner}W. Greiner, {\it Relativistic Quantum Mechanics}, (Springer, Heidelberg, 1990). The Dirac equation for the Coulomb central potential is discussed on page 178.
\bibitem{WS} R. D. Woods and D. S. Saxon, Phys. Rev. {\bf 95}, 577 (1954).
\bibitem{Yuk} H. Yukawa, Proc. Phys. Math. Soc. Japan. {\bf 17}, 48 (1935).
\bibitem{Hult} L. Hulthen, Ark. Mat. Astron. Fys. {\bf 28 A}, 5 (1942); {\it ibid}., Ark. Mat. Astron. Fys. {\bf 29 B}, 1 (1942).
\bibitem{C-E} A. Jeffrey, {\it Advanced engineering mathematics}, (Academic Press, Burlington, 2002). Cauchy-Euler equation is disscussed on page 309.
\bibitem{PT} G. P\"{o}schl, E. Teller, Z. Phys. {\bf 83}, 143 (1933).
\bibitem{Eckart} C. Eckart, Phys. Rev. {\bf 35}, 1303 (1930).
\end{thebibliography}
\end{document}